\documentclass[aps,prx,twocolumn,superscriptaddress]{revtex4-2}
\usepackage[utf8]{inputenc}

\usepackage[titletoc,toc,title,page]{appendix}

\usepackage{csquotes,graphicx}
\usepackage[italian,english]{babel}

\usepackage{microtype} 

\usepackage{bm} 

\usepackage{dsfont} 
\usepackage{amsmath,amssymb,amsthm,thmtools}
\usepackage{mathtools}
\usepackage{cases}
\usepackage{calc,palatino}
\usepackage{mathrsfs} 
\usepackage[normalem]{ulem} 

\usepackage{parskip}
\usepackage{nameref}
\usepackage[colorlinks=true]{hyperref}
\usepackage[nameinlink]{cleveref}
\crefname{appsec}{Appendix}{Appendices}
\crefname{box}{Box}{Box}
\hypersetup{
  colorlinks   = true, 
  urlcolor     = green!80!black, 
  linkcolor    = blue, 
  citecolor    = red!80!black 
}

\usepackage{physics} 

\usepackage{float} 

\usepackage{graphicx}

\usepackage[usenames,dvipsnames,table]{xcolor}
\usepackage{easyReview}

\usepackage{tikz}
\usetikzlibrary{calc,shapes.geometric}

\usepackage{placeins}
\usepackage{multirow,tabularx,booktabs}
\usepackage[most]{tcolorbox}
\newtcbtheorem{tbox}{Box}{enhanced, float*=t, width=\textwidth, label type=box}{box}

\usepackage[printonlyused,withpage,nohyperlinks,smaller]{acronym}

\graphicspath{{./figures/}}

\newcommand{\calD}{\mathcal{D}}

\newcommand{\calL}{\mathcal{L}}

\begin{document}

\title{Non-equilibrium thermodynamics of gravitational objective-collapse models}
\author{Simone Artini}
\let\comma,
\affiliation{Quantum Theory Group\comma{} Dipartimento di Fisica e Chimica Emilio Segr\`e\comma{} Universit\`a degli Studi di Palermo\comma{} via Archirafi 36\comma{} I-90123 Palermo\comma{} Italy}
\author{Gabriele Lo Monaco}
\let\comma,
\affiliation{Quantum Theory Group\comma{} Dipartimento di Fisica e Chimica Emilio Segr\`e\comma{} Universit\`a degli Studi di Palermo\comma{} via Archirafi 36\comma{} I-90123 Palermo\comma{} Italy}
\author{Sandro Donadi}
\affiliation{Centre for Quantum Materials and Technologies\comma{} School of Mathematics and Physics\comma{} Queen’s University Belfast\comma{} BT7 1NN\comma{} United Kingdom}
\affiliation{Istituto Nazionale di Fisica Nucleare\comma{} Trieste Section\comma{} Via Valerio 2\comma{} 34127 Trieste\comma{} Italy}
\author{Mauro Paternostro}
\let\comma,
\affiliation{Quantum Theory Group\comma{} Dipartimento di Fisica e Chimica Emilio Segr\`e\comma{} Universit\`a degli Studi di Palermo\comma{} via Archirafi 36\comma{} I-90123 Palermo\comma{} Italy}
\affiliation{Centre for Quantum Materials and Technologies\comma{} School of Mathematics and Physics\comma{} Queen’s University Belfast\comma{} BT7 1NN\comma{} United Kingdom}


\begin{abstract}
We investigate the entropy production in the Di\'osi-Penrose (DP) model, one of the most extensively studied gravity-related collapse mechanisms, and one of its  dissipative extensions. 
To this end, we analyze the behavior of a single harmonic oscillator, subjected to such collapse mechanisms, focusing on its phase-space dynamics and the time evolution of the entropy production rate — a central quantity in non-equilibrium thermodynamics. Our findings reveal that the original DP model induces unbounded heating, producing dynamics consistent with the Second Law of thermodynamics only under the assumption of an infinite-temperature noise field. In contrast, its dissipative extension achieves physically consistent thermalization in the regime of low dissipation strength. We further our study to address the complete dynamics of the dissipative extension, thus including explicitly non-Gaussian features in the state of the system that lack from the low-dissipation regime, using a short-time approach. 

\end{abstract}

\maketitle

\section{Introduction}

One of the most important open problems in theoretical physics is how to reconcile gravity and quantum mechanics. Both theories work extremely well within their respective domains of validity, but finding a unified description from which both can be derived remains a profound challenge. 
{It is largely believed that gravity, like other fundamental interactions, should manifest typical quantum effects at short distances and  it should thus be  quantized.} Despite extensive efforts in this direction, a fully {coherent} quantum theory of gravity {whose predictions can be tested and the currently accessible scales of energies} remains elusive.

{Different approaches renounce to find a proper quantization procedure for gravity and rather propose effective descriptions of how gravitational interaction may influence the dynamics of quantum mechanical systems. One of the possible realizations of this program is the idea by Penrose of ``gravitizing" quantum theory by introducing a spontaneous collapse of the wave function induced by the classical gravitational interaction \cite{penrose1996gravity}.} In this way, a new approach to merging gravity and quantum theory emerges, which not only attempts to reconcile the two but also provides a solution to the quantum measurement problem \cite{bassi2013models}.

A first pioneering step towards such models was made by Károlyházy, who suggested that gravitational effects could induce quantum decoherence \cite{karolyhazy1966gravitation}. Later, the first proper collapse model that relates spontaneous wave function collapse to gravity was introduced by Di\'osi \cite{diosi1987universal,diosi1989models}. The timescale for the collapse of a spatial superposition predicted by this model coincides with that later proposed by Penrose \cite{penrose1996gravity}, and for this reason, the two proposals are nowadays commonly referred to as the Di\'osi-Penrose (DP) model. In the last 30 years, several models predicting quantum decoherence related to gravity have been put forward \cite{hu2008stochastic,breuer2009metric,blencowe2013effective,gasbarri2017gravity,asprea2021gravitational} (see \cite{bassi2017gravitational} for a more complete review of existing models). Gravitational wave-function collapse and decoherence are also predicted by models that treat gravity as fundamentally classical \cite{diosi1998coupling,diosi2000quantum,diosi2023hybrid,oppenheim2023postquantum}. 

In this work, in order to address a reference model that is both simple and significant, we focus on the DP model and its dissipative generalizations, which  modify the Schr\"odinger equation by introducing additional terms {responsible for the wave function to collapse in space} \cite{diosi1987universal,diosi1989models}. It has been shown that the induction of collapse mechanisms in the dynamics comes with an associated diffusive behavior \cite{donadi2023collapse}. The latter has been exploited in experiments involving bulk matter \cite{vinante2021gravity} and radiation emission from germanium \cite{donadi2021underground,piscicchia2024x} to provide stringent bounds on the free parameter  of the model \cite{arnquist2022search}.

In the standard DP model, such diffusion also manifests itself as a steady increase in energy of the center of mass, which occurs even for isolated systems. This violates energy conservation, albeit at a level too small to be detected in current experiments for a large range of values of the free parameter $R_0$. While violations of energy conservation may be tolerable from a phenomenological perspective, the nonphysical nature of a persistent energy increase is problematic even within the framework of an effective model; one would expect a system to thermalize with the noise that induces collapse, rather than exhibit an unbounded energy growth.

To address this issue, two dissipative extensions of the DP model have been proposed \cite{smirne2015dissipative,di2023linear}. These models introduce a mechanism that allows the system to reach thermal equilibrium with the collapse-inducing noise, hence mitigating the energy-growth problem and leading to more consistent dynamics from a thermodynamical point of view.  
 {In Ref.~\cite{artini2023characterizing}, violations of the Second Law of thermodynamics have been observed also in the Continuous Spontaneous Localization (CSL) \cite{ghirardi1990markov}, which is another model designed to solve the quantum measurement problem. Here, we closely scrutinize the thermodynamic consistency of the DP model and its dissipative generalization introduced in Ref.~\cite{di2023linear}.} We show that the dissipative extension of the DP model, for small values of the parameter that controls the strength of the friction, adheres to the Clausius law of thermodynamics, leading to a physically consistent thermal equilibration, in contrast to the original DP model in which thermal equilibrium is never reached and no violations of the Second law are witnessed only for an infinite temperature noise field. However, considering the full non-gaussian evolution stemming from the dissipative DP model, clues of such violations are still found at very early times for low temperatures.
 To reach these conclusions, we study the entropy production rate \cite{landi2021irreversible,santos2018irreversibility,deffner2011nonequilibrium} of the dynamics, a quantity intimately related to the Second Law, over time exploiting the phase-space picture of the dynamics \cite{man2002alternative,santos2017wigner}.

The paper is organized as follows: in Sec.~\ref{sec: Entropy production rate and equilibration processes} we review basic concepts of thermodynamics, both in and out equilibrium, and we introduce the entropy production rate, {the main tool} used in the remainder of the paper. In Sec.~\ref{sec: Frictionless Diosi-Penrose model} we compute the equivalent Fokker-Planck equation -- which describes the dynamics of the system in phase space in terms of quasiprobability distributions -- 
 for the standard DP model and study the entropy production rate of a harmonic oscillator subject to such dynamics. In Sec.~\ref{sec: Linear-friction Diosi-Penrose model} we further the analysis to the dissipative extension of the DP model introduced in Ref. \cite{di2023linear}, but focusing on the case of small dissipation strength. Sec.~\ref{sec: non gauss} is dedicated to the analysis of the complete dynamics, which involve non-Gaussian terms in the evolution. The analysis is carried out using a technique that linearizes the Lindbladian for small times, similar to the linear-response method used in  Refs.~\cite{PhysRevD.111.026009,Konopik2019, Blair2024}. We find inconsistencies with the Second Law of Thermodynamics, which, however, may be due to the approximations made.; finally, in Section \ref{sec: concl} we summarize and discuss the findings of the analysis.

\section{Entropy production rate and equilibration processes}
\label{sec: Entropy production rate and equilibration processes}
Unlike energy, the entropy of an open system  does not satisfy a continuity equation: its rate of change depends on entropic fluxes exchanged with the environment with which the system interacts and a process of entropy
production that is intrinsic to the system itself. According to the Second Law
of thermodynamics, the latter contribution is always non-negative and null only
when the system and the environment are in thermal equilibrium. 
 
By calling $\Sigma$ the entropy production and $\Phi$ the entropy flux associated with a given open dynamics, the change of entropy of the system is ruled by the following  reformulation of Clausius' theorem of {\it uncompensated heat}
\begin{equation}
\Delta\text{S}_{\text{sys}} = \Sigma - \Phi.
\end{equation}
In the quasi-static limit where the system goes across a series of equilibrium states, we have $\Sigma = 0$, so that the change of entropy of the system is entirely ascribed to the entropy flux. The entropy production thus  serves as a measure of the irreversibility
of a physical process and may be used to characterize non-
equilibrium systems in a broad range of situations and acrossall length scales~\cite{landi2021irreversible}. 
It is often convenient to consider the differential form of Clausius' theorem $\dot{\text{S}}_{sys}= \Pi(t) - \phi(t)$, where we have introduced the {\it entropy production rate} $\Pi(t) = \dot{\Sigma}$ and {\it entropy flux rate} $\phi(t)= \dot{\Phi}(t)$. 
The entropy production rate, a key quantity in the study of out-of-equilibrium processes, must satisfy $\Pi(t)\geq 0\, , \, \forall t$ for consistency with the Second Law of thermodynamics, as it represents the rate of the total entropy {change}. 

Our aim is to study a quantum system that is subject to a collapse dynamics, which can be effectively described by a noise field induced by a thermal bath at a certain (but not necessarily finite) temperature and to characterize its dynamics using the quantities introduced above to assess its physical consistency from the thermodynamic point of view. As our fundamental figure of merit for entropy production, we use the Rényi$-2$ entropy $S_2 = -\ln({\rm Tr}\hat{\rho}^2)$, which has the advantage of being well-behaved in situations where the von Neumann entropy is pathological (the so-called ultra-cold catastrophe)~\cite{uzdin2021passivity}. For states with a poisitive Wigner function, the Rényi$-2$ entropy is particularly handy as it is equivalent to the Wigner entropy $S_W=-\int W(q,p)\ln W(q,p)$ with $W(q,p)$  the Wigner quasiprobability distribution. Assuming such figure of merit, one can define the associated entropy production rate  
\begin{equation}
\label{def: ent prod rate}
\begin{aligned}
    \Pi(t)&= -\partial_t K(W(t)||W_{eq})\\
    &:= -\partial_t \int dq\, dp\, W(q,p,t)\ln\frac{W(q,p,t)}{W_{eq}(q,p)},
\end{aligned}
    \end{equation}
where $W(t)$ is the Wigner function of the system at time $t$, $W_{eq}$ is that of a suitable target state -- usually the asymptotic equilibrium state of the dynamics~\cite{santos2017wigner} -- and $K(p||q) = \int dx dy p(x,y)\ln(p(x,y)/q(x,y))$ is the Kulback-Leibler divergence of the probability distributions $p(x,y)$ and $q(x,y)$ \cite{cover1991information}. Notice that $\Pi(t)=0$ if and only if $W(t)=W_{eq}$ i.e. when the system is at equilibrium (provided that the stationary state is unique). 

In what follows, we work using rescaled position and momentum operators that obey the commutation relations $[\hat q_i,\hat p_j]=2i\hbar\delta_{ij} $, with $i,j=x,y,z$ labelling different components of the corresponding operators.
\section{Frictionless Di\'osi-Penrose model}
\label{sec: Frictionless Diosi-Penrose model}

In this section, we focus on the simplest version of the DP model \cite{diosi1987universal,diosi1989models,penrose1996gravity,diosi2007notes}, which we dub \emph{frictionless}. In order to fix the ideas without unnecessary complications, we also restrict our attention to the case of a single particle. 
In the DP model, the evolution of the statistical operator $\hat \rho(t)$ of a particle is given by the Lindblad master equation
\begin{equation}
\label{eq:lindblad}
    \partial_t \hat \rho(t)=-\frac{i}{\hbar}[\hat H,\hat \rho(t)]+\mathcal{L}[\hat \rho(t)]
\end{equation}
with $\hat{H}$ being the Hamiltonian describing the system.
The superoperator $\mathcal{L}$ is the one responsible for the gravitational (dynamical) collapse and it is written as \begin{equation}
\label{eq:L}
    \mathcal{L}[\hat \rho]=\frac{1}{\hbar^2}\int \frac{d^3k}{(2\pi)^3}\Gamma(k)\left(\hat L_k\hat \rho \hat L^\dagger_k-\frac{1}{2}\left\{\hat L^\dagger_k\hat L_k,\hat \rho \right\}\right)\,,
\end{equation}

with $\hat{L}_k=me^{ik\cdot\hat{x}}$ and $\Gamma_{\rm DP}(k)={4\pi\hbar G}e^{-k^2 R_0^2}/k^2$. Here, $R_0$ is a cutoff with the dimensions of a length. The lower bounds on $R_0$ come from comparison with experiments, with the stringent lower bound $R_0\geq 4\times 10^{-10}$ m set by experiments measuring spontaneous radiation emission from Germanium \cite{donadi2021underground, arnquist2022search}. Upper bounds on $R_0$ are set by theoretical arguments, requiring that the collapse should be strong enough to guarantee that macroscopic objects are well localized, leading to 
 $R_0\lesssim 10^{-4}$ m \cite{figurato2024effectiveness}.

\subsection{Fokker-Planck equation}
\label{subsec: Frictionless Fokker-Planck equation}
Our choice of thermodynamic figure of merit for the quantification of the irreversible entropy will require to translate the collapse dynamics to the phase-space picture via a Wigner-Weyl transform~\cite{man2002alternative}. 
{As we will show soon, the Lindblad equation describing the evolution of the statistical operator translates, under appropriate assumptions, into a Fokker-Planck equation for the Wigner function in the phase space~\cite{artini2023characterizing}.}
For a generic operator $\hat A$, the Wigner function is defined as
\begin{equation}
\label{eq:wigner_position}
    W_{\hat A}(\bm q, \bm p)=\int \frac{d^3y}{(4\pi)^3}e^{-i\frac{\bm p\cdot \bm y}{2\hbar}}\bra{\bm q+\bm y/2}\hat A\ket{\bm q-\bm y/2}.
\end{equation}
{To handle the commutator on the right-hanf side of} \cref{eq:lindblad}, we exploit the {well-known} identity $W_{[\hat A,\hat B]}=\{W_{\hat A},W_{\hat B}\}_*$~\cite{baker1958formulation}, {where}
\begin{equation}
\label{der}
    \{W_{\hat A},W_{\hat B}\}_*=2 W_{\hat A}\sin(\frac{1}{2}\Omega^{IJ}\overset{\leftarrow}{\partial_I}\overset{\rightarrow}{\partial_J})W_{\hat{B}}
\end{equation}
{is the} Moyal bracket. In \cref{der}, {the indices}  $I,J$ {run} over the phase space coordinates $\bm X=(\bm q,\bm p)$ and $\Omega=\left(\begin{matrix} 0 & \mathbb{I}_3\\ -\mathbb{I}_3 & 0\end{matrix}\right)$ is the symplectic matrix. As for the Wigner term corresponding to the anti-commutator term in the Lindbladian, this can be 
straightforwardly computed as $-\Lambda_{\rm DP} W_{\hat \rho(t)}$,  where we have introduced the rate 
\begin{equation}
    \Lambda_{\rm DP}=\frac{m^2}{\hbar^2}\int \frac{d^3k}{(2\pi)^3}\,\Gamma_{\rm DP}(k)=\frac{Gm^2}{\sqrt{\pi}\hbar R_0}\,.
\end{equation}
We thus remain with the problem of calculating the Wigner term corresponding with $\hat L_k\hat \rho \hat L^\dagger_k$, that is
\begin{equation}
\frac{m^2}{\hbar^2}\int \frac{d^3k}{(2\pi)^3}\,\Gamma_{\rm DP}(k)\,W_{\hat \rho}(\bm q,\bm p-2\hbar k).
\end{equation}
The dependence of $W_{\hat\rho(r)}$ on ${\bm p}-2\hbar k$ is a consequence of the specific form of the commutation relations that we have adopted, and we have omitted the time dependence of the density matrix for simplicity of notation. This expression can be further elaborated by assuming well-localized states and expanding the Wigner function as

\begin{equation}
\begin{aligned}
   &\frac{m^2}{\hbar^2}\int\!\! \frac{d^3k}{(2\pi)^3}\,\Gamma_{\rm DP}(k)\,W_{\hat \rho}(\bm q,\bm p-2\hbar k)=\!  \\
   &=\!\!\frac{m^2}{\hbar^2}\int\!\! \frac{d^3k}{(2\pi)^3}\,\Gamma_{\rm DP}(k)\,\left[1-2\hbar k\cdot\nabla_{\bm p}+2(\hbar k\cdot\nabla_{\bm p})^2\right]W_{\hat \rho}\,\!\nonumber\\
   &=\!\Lambda_{\rm DP}W_{\hat \rho}+\!\frac{2m^2}{\hbar^2}\int\!\! \frac{d^3k}{(2\pi)^3}\,\Gamma_{\rm DP}(k)(\hbar k\cdot \nabla_{\bm p})^2W_{\hat \rho},
\end{aligned}
\end{equation}

where we used the fact that $\Gamma_{\rm DP}(k)$ is an even function of $k$. 
By symmetry, we also observe that 
\begin{equation}
\begin{split}
    & \frac{m^2}{\hbar^2}\int \frac{d^3k}{(2\pi)^3}\,\Gamma_{\rm DP}(k)(\hbar k\cdot \nabla_{\bm p})^2W_{\hat \rho}\,=\,
   \frac{\hbar Gm^2}{6\sqrt{\pi} R_0^3}\Delta_{\bm p}W_{\hat \rho}.
\end{split}
\end{equation}
with $\Delta_{\bm p}=\partial^2_{p_x}+\partial^2_{p_y}+\partial^2_{p_z}$. Collecting all the terms together, we get the  Fokker-Planck equation
\begin{equation}
\label{DP_FP}
    \partial_tW_{\hat\rho(t)}\,=\,\{W_{\hat H},W_{\hat \rho(t)}\}_\star+D\,\Delta_{(\bm p)}W_{\hat \rho(t)}
\end{equation}
with $D=\frac{G\hbar m^2}{3\sqrt{\pi} R_0^3}$ the diffusion parameter. 
\Cref{DP_FP} can be compared with the Fokker-Planck equation obtained for the mass-proportional version of the CSL model in three dimensions~\cite{artini2023characterizing}, whose Lindblad operator reads as in ~\cref{eq:L} but with $\Gamma_\text{DP}(k)\to\Gamma_{\rm CSL}(k)=\gamma \hbar^2\bar{m}^2\exp[-\frac{k^2 r_c^2}{\hbar^2}]/(8m^2\pi^3)$, where $\bar{m}$ is a dimensionless mass written in units of the reference mass $m_0=1$ amu~\cite{bahrami2014role}. Within the domain of validity of the approximation used in the previous calculation for Wigner functions concentrated around the origin, we get to
\begin{equation}
\begin{split}
     W_{\mathcal{L}[\hat \rho]} &= 2\frac{m^2}{\hbar^2}\int \frac{d^3k}{(2\pi)^3}\,\Gamma_{\rm CSL}(k)\nabla_{\bm p}^2W_{\hat \rho}\\
    &=\frac{\gamma\,m^2}{8\sqrt{\pi^3}m_0^2r_c^5}\Delta_{\bm p}W_{\hat \rho}
\end{split}
\end{equation}
\medskip
where we have used simple Gaussian integration. We can switch to the usual CSL formulation using $\gamma = 8\pi^{3/2}\lambda r_c^3$, which yields $D_{\rm CSL} =  \frac{\lambda}{r_c^2} \frac{m^2}{m_0^2}$. The ratio between the two diffusion parameters coincides with the ratio of the heating powers of the two models as computed in Ref.~\cite{di2023linear}, suggesting the existence of a mapping between the two models (at least for the case of a single particle).

\subsection{Solution and entropy dynamics}
\label{subsec: Solution and entropy dynamics 1}
We can now numerically solve~\cref{DP_FP} and study the entropy production rate over time of a system subject to the collapse dynamics. We start by noting that for a factorized initial state, solutions of \cref{DP_FP} are factorized as well. In this case, one can focus in the dynamics in one dimension, which simplifies the numerical evaluation. From the thermodynamic point of view, the three spatial degrees of freedom will indeed give independent contributions, thus making a one-dimensional study fully fit for the purpose of gathering an understanding of the energetics of the model at hand~\footnote{Consider the Shannon entropy of a factorized probability distribution: $p(x_1,x_2,x_3)=p_1(x_1)p_2(x_2)p_3(x_3)$, it is easy to show that $S(p) = S(p_1)+S(p_2)+S(p_3)$. The same holds for the relative entropy as long as the reference distribution is factorized as well.}. 

\begin{figure} [t!]

{\bf (a)}
    \includegraphics[width=\columnwidth]{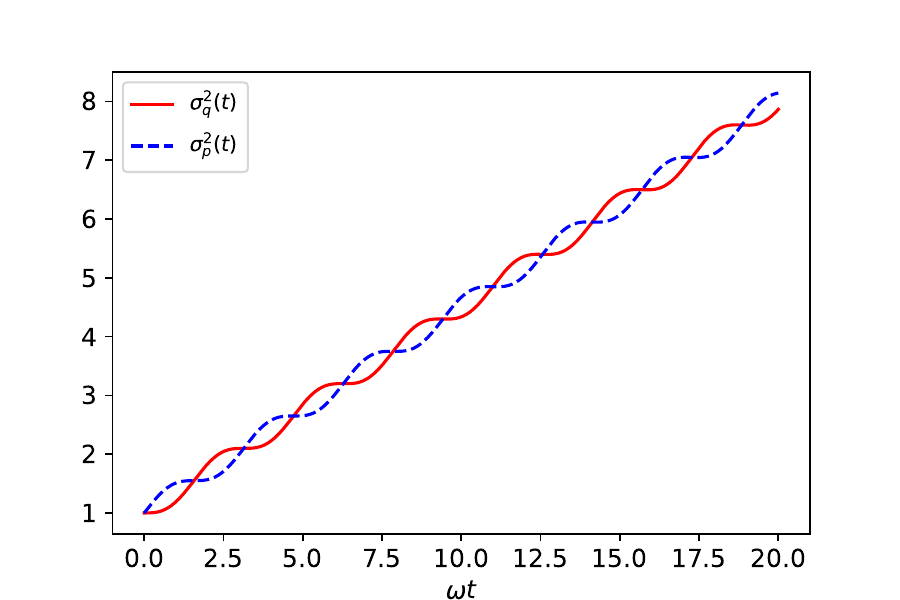}

{\bf (b)}
    \includegraphics[width=\columnwidth]{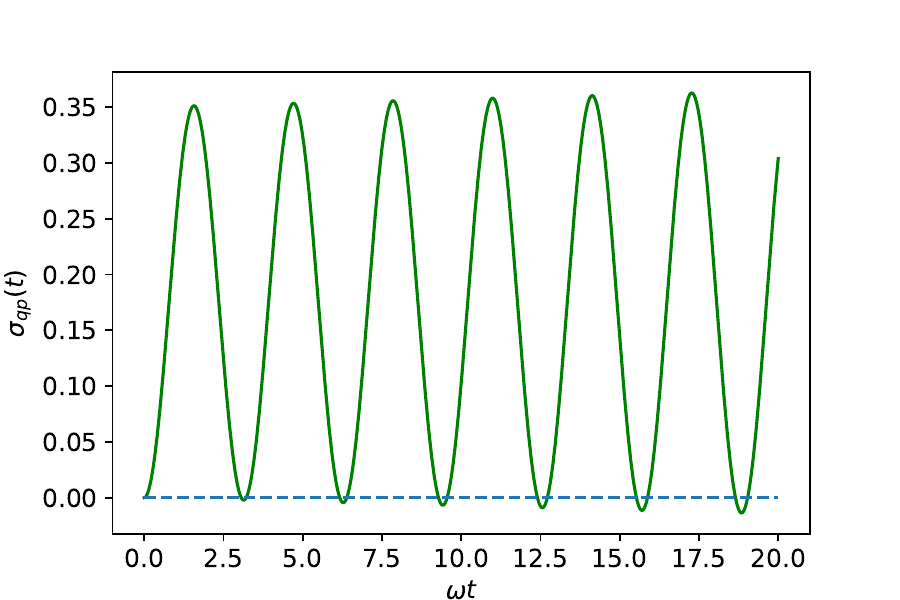}
 
    \caption{Evolution of the covariance matrix entries over time. We have integrated the corresponding equations of motion by discretizing time using $N=10^4$ time steps. {\bf (a)}: We show the position variance $\sigma_q^2(t)$ (solid red line) and momentum variance $\sigma_p^2(t)$ (in (dashed blue line) against time within the interval $T=20$. {\bf (b)}: We display the behaviour of the covariance $\sigma_{pq}(t)=\langle\{\hat q,\hat p\}\rangle$. The initial state is the ground state of the oscillator. The state is heating indefinitely oscillating around the linear increase in variance of a diffusive dynamics. All quantities reported are dimensionless and we have rescaled the parameters so that $\hbar=G=1$.}
\label{fig:variances_fless}
\end{figure}

For the case study, we consider an harmonic oscillator with frequency $\omega$ initially prepared in the thermal state at inverse temperature $\beta_o$, whose Wigner function is $W_\text{Th} (x)= \dfrac{e^{-\frac{1}{2}x^TV_{Th}^{-1}x}}{2\pi(2\bar{n}+1)}$ with $x=(q,p)$the phase-space coordinates, $V_{Th} = (2\bar{n}+1)\mathbb{I}$ the covariance matrix (here $\mathbb{I}$ is the $2\times2$ identity matrix), and $\bar{n}=\left(e^{\beta_o\hbar\omega}-1\right)^{-1}$ the thermal occupation number. The quadratic nature of the dynamics allows to track the properties of the  Wigner function of the oscillator simply through the knowledge of the temporal behavior of its covariance-matrix elements in the form of a set of coupled ordinary differential equations. In order to solve the equations of motion for such entries, we set $\hbar = G = 1$ and integrate over dimensionless time $\omega t$ with respect to the frequency of the oscillator for a time interval $T=20$ with $N=10^4$ time steps. The mass of the oscillator, $m$, and the parameter of the DP model, $R_0$, determine the strength of the diffusion through the parameter $D$.  For clarity in visualization and to ensure readable plots we selected the numerical values $m=2$ and $R_0=3$, being interested only in the overall features of the dynamics. The results for the entries of the covariance matrix, starting from the ground state of the oscillator, are reported in Fig.~\ref{fig:variances_fless}. The variances of both momentum and position grow oscillating -- the oscillations being due to the Hamiltonian part of the evolution, resulting in a   rotation in phase-space -- around an otherwise linear trend, which is due to the purely diffusive part of the dynamics. This entails that they never reach an asymptotic value. The absence of an asymptotic solution can be indeed surmised from the equations of motion of the covariance matrix entries as well~\cite{artini2023characterizing}.

\begin{figure*}[t!]
{\bf (a)}\hskip9cm{\bf (b)}\\
    \includegraphics[width=\columnwidth]{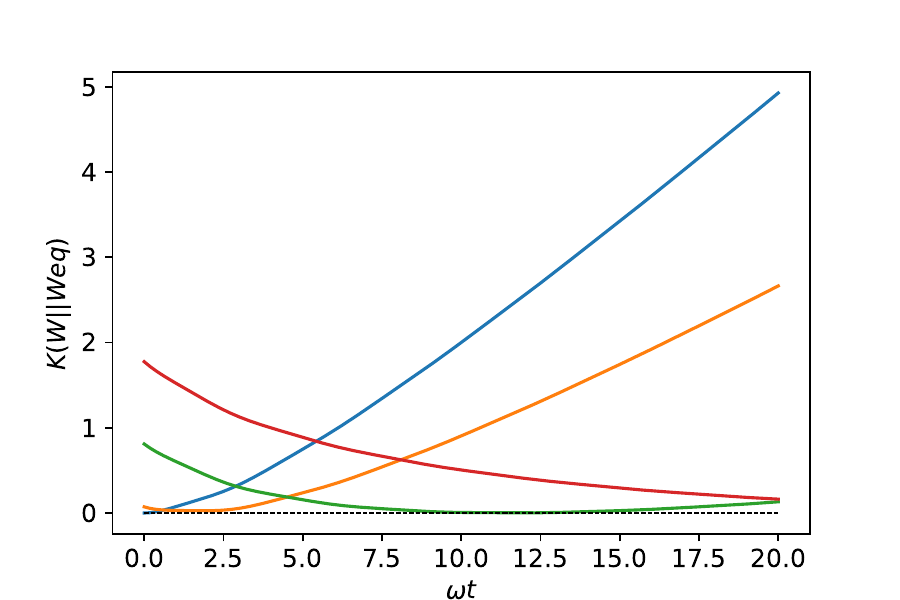}    \includegraphics[width=\columnwidth]{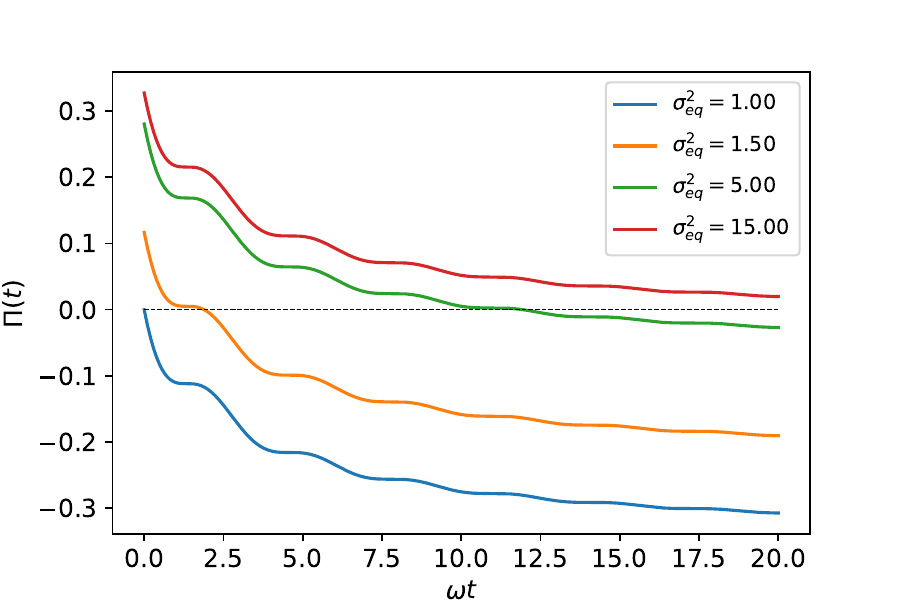}
    \caption{{\bf (a)}: Relative entropy between the state of the system at time $t$ and target states of increasing temperatures. {\bf (b)}: Corresponding entropy production rate. The relative entropy becomes null at finite time, becoming then negative. This can be avoided only by assuming an infinite temperature target state. All parameters as in Fig.~\ref{fig:variances_fless}.}
\label{fig:entropy_fless}
\end{figure*}

Using the entropy production rate as the figure of merit computed as defined in \cref{def: ent prod rate}, it is possible to show that this dynamic must be the result of the system being in contact with an infinite temperature bath, never reaching equilibrium. In Fig.~\ref{fig:entropy_fless} the relative entropy and the entropy production rate are reported using the same setup of Fig.~\ref{fig:variances_fless} assuming as the target state for the relative entropy a thermal state with increasing temperature. Indeed, one can see that, for any finite temperature, the entropy production rate decreases from a finite value, reaches zero at a certain time and then becomes negative violating the Second Law. Increasing the target temperature delays the time at which the violation happens, suggesting that the target temperature should be infinite in order to have a physically consistent dynamics. Furthermore, it is evident that, in such case ,the system remains out-of-equilibrium indefinitely since $\Pi \to 0$ only for $t \to +\infty$.

\section{Linear-friction Di\'osi-Penrose model}
\label{sec: Linear-friction Diosi-Penrose model}
Due of {the} undesired heat generation in the original Di\'osi-Penrose model, it has been proposed to introduce a dissipation mechanism in the model \cite{bahrami2014role,di2023linear}. In this section, we focus on the \emph{linear-friction} model of Ref.~\cite{di2023linear}.
For a single particle of mass $m$, the new Lindbladian is as in \cref{eq:L} but with the following modified Lindblad operators:
\begin{align}
\hat L_k&=me^{ik\cdot \hat x}-\frac{\hbar \beta}{8}\left\{k\cdot \hat p,e^{ik\cdot \hat x}\right\}\\
&=\left(m+\frac{\hbar^2\beta}{4}k^2-\frac{\hbar \beta}{4}k\cdot \hat p\right)e^{ik\cdot \hat x}\nonumber\\
&=e^{ik\cdot \hat x}\left(m-\frac{\hbar^2\beta}{4}k^2-\frac{\hbar \beta}{4}k\cdot \hat p\right)\,.\nonumber
\end{align}
The function $\Gamma(k)$ depends on the specific collapse model
being addressed as
\begin{equation}
\label{eq: Gamma}
    \Gamma(k)=e^{-k^2R_0^2} \left\{\begin{matrix}
        4\pi \hbar G/k^2 & \quad ({\rm DP}), \\
        \hbar^2 \gamma & \quad ({\rm CSL}). 
    \end{matrix}\right.
\end{equation}
The parameter $\beta$ quantifies the strength of the dissipation and is related [in a non-trivial way, cf.~\cref{eq: temp}] to the inverse temperature of the stationary Gibbs state of the dynamics.

\subsection{Fokker-Planck equation}
\label{subsec: Fokker-Planck equation}
{The frictional case} {can also be tackled by casting the dynamics in terms of} a {higher-derivative} Fokker-Planck equation for {the} Wigner function associated with the state of the system. To this goal, it is convenient to work with the momentum-space representation of the Wigner function (cf.~\cref{app:wigner_momentum})
\begin{equation}
\label{eq:wigner_momentum}
    W_{\hat A}(q,p)=\int\frac{d^3 v}{(4\pi)^3}e^{-i\frac{q\cdot v}{2\hbar}}\bra{p-\frac{v}{2}}\hat A \ket{p+\frac{v}{2}}
\end{equation}
with $\ket{p}$ being momentum eigenstates.  
We start observing that 
\begin{widetext}
\begin{equation}
\begin{aligned}
&\int \frac{d^3v\,e^{-i \frac{q\cdot v}{2\hbar}}}{(4\pi)^3}\bra{p-\frac{v}{2}}\hat L_k\hat \rho \hat L^\dagger_k\ket{p+\frac{v}{2}}= \int \frac{d^3v\,e^{-i \frac{q\cdot v}{2\hbar}}}{(4\pi)^3}\bra{p-2\hbar k-\frac{v}{2}}\hat \rho \ket{p-2\hbar k+\frac{v}{2}}\left(m_{(+)}-\frac{\hbar \beta}{8}k\cdot v\right)\!\left(m_{(+)}+\frac{\hbar \beta}{8}k\cdot v\right),
\end{aligned}
\end{equation}
\end{widetext}
where we have introduced
\begin{equation}
     m_{(\pm)}=m\pm \frac{\hbar^2\beta}{4}k^2-\frac{\hbar \beta}{4}k\cdot p
\end{equation}
and used $e^{ik\cdot \hat x}\ket{p}=\ket{p+2\hbar k}$. Using the effective replacement $v\to2i\hbar \partial_q$ in the integral, we arrive at
\begin{equation}
\begin{aligned}
\int&\frac{d^3v}{(4\pi)^3}e^{-i \frac{q\cdot v}{2\hbar}}\bra{p-\frac{v}{2}}\hat L_k\hat \rho \hat L^\dagger_k\ket{p+\frac{v}{2}}\\
&=\left[m_{(+)}^2+\frac{\hbar^4\beta^2}{16}(k\cdot \partial_q)^2\right]W_{\hat \rho}(q,p-2\hbar k).
\end{aligned}
\end{equation}
Similarly
\begin{equation}
\begin{aligned}
\frac{1}{2}\int&\frac{d^3v}{(4\pi)^3}e^{-i \frac{q\cdot v}{2\hbar}}\bra{p-\frac{v}{2}}\left\{\hat L^\dagger_k\hat L_k,\hat \rho \right\}\ket{p+\frac{v}{2}}=\\
&=\left[m_{(-)}^2-\frac{\hbar^4\beta^2}{16}(k\cdot \partial_q)^2\right]W_{\hat \rho}(q,p).
\end{aligned}
\end{equation}
 We focus on the regime where $W(q,p-2\hbar k)$ can be expanded in power series of $k$, truncating it at the second order. The final differential equation for the Wigner function thus read
\begin{equation}\label{full_W}
\begin{aligned}
    \partial_t W_{\hat \rho}&=\left\{W_{\hat H},W_{\hat \rho}\right\}_\star+\nabla_{\bm p}\!\cdot\!(f\bm p W_{\hat \rho})+\partial_{p_i}\partial_{p_j}(D_{ij}W_{\hat \rho})\\
    &+\partial_{q_i}\partial_{q_j}(\tilde D_{ij}W_{\hat \rho})+ R^{ijkl}\partial_{p_i}\partial_{p_j}\partial_{q_j}\partial_{q_k}W_{\hat \rho}.
\end{aligned}
\end{equation}
Let us address the various terms entering this equation. We start remarking that the fourth-order derivative appearing in \cref{full_W} is {typical of this model}. For instance, similar corrections to the standard Fokker-Planck equation are quite common in the context of hydrodynamics and diffusion processes, a typical example being the Swift-Hohenberg equation \cite{swift1977hydrodynamic}. The drift vector of the form $\mu_i=f p_i$ involves the scalar term
\begin{equation}
    f= \frac{m \beta}{3}\int \frac{d^3k}{(2\pi)^3}k^2\Gamma(k)\left(1-\frac{\hbar^2 \beta}{4m}k^2\right).
\end{equation}
Diffusion processes involve a momentum term characterized by the tensor
\begin{equation}
D_{ij}=D_1\delta_{ij}+D_2 p^2 \delta_{ij}+D_3\,p_i p_j
\end{equation}
with
\begin{equation}
\begin{aligned}
    D_1&=\frac{2 m^2}{3}\int\frac{d^3k}{(2\pi)^3}k^2\Gamma(k)\left(1+\frac{\hbar^2\beta}{4m}k^2\right)^2,\\
    D_3&=2D_2=\frac{\hbar^2 \beta^2}{60}\int \frac{d^3k}{(2\pi)^3}k^4 \Gamma(k).
\end{aligned}
\end{equation}
As for the position-related part, the tensor $\tilde D_{ij}$ takes the form
\begin{equation}
    \tilde D_{ij}\,=\,\delta_{ij}\frac{\hbar^2\beta^2}{24}\int \frac{d^3k}{(2\pi)^3}k^2\Gamma(k).
\end{equation}
Finally
\begin{equation}
R^{ijkl}=\frac{\hbar^2\beta^2}{120}(\delta^{ij}\delta^{kl}+\delta^{ik}\delta^{jl}+\delta^{il}\delta^{jk})\int \frac{d^3k}{(2\pi)^3}k^4 \Gamma(k).
\end{equation}
Assuming the dissipation parameter $\beta$ to be small, \cref{full_W} reduces to a functional form that is analogous to the standard Klein-Kramers equation that describes classical Brownian motion \cite{kramers1940brownian}
\begin{equation}
\label{eq: smallBetaQFP}
    \partial_t W_{\hat \rho}=\left\{W_{\hat H},W_{\hat \rho}\right\}_\star+\nabla_{\bm p}\cdot(f\,\bm p \,W_{\hat \rho})+D\Delta_{\bm p}W_{\hat \rho}\,,
\end{equation}
with 
\begin{equation}
\begin{aligned}
    \label{D}
    D&=\frac{2m^2}{3}\int\frac{d^3k}{(2\pi)^3}k^2\Gamma(k)\left(1+\frac{\hbar^2\beta}{2m}k^2\right),\\
    f&=\frac{m \beta}{3}\int \frac{d^3k}{(2\pi)^3}k^2\Gamma(k)\,.
    \end{aligned}
\end{equation}
In this conditions, thus, diffusion will depend on the tensor $D_{ij}=D\delta_{ij}$, while the drift vector will read $\mu_i=f p_i\,$.
In the frictionless limit $\beta \rightarrow 0$, the diffusion coefficient coincides with the analogous one for the frictionless CSL/DP model.

\subsection{Solution and non-negativity of entropy production rate in the small $\beta$ regime}
\label{subsec: Solution and entropy dynamics 2}
In the limit of small values of $\beta$, the form of \cref{eq: smallBetaQFP} allows to prove the non-negativity of entropy production rate by following the same arguments of Ref. \cite{santos2017wigner} or  Ref. \cite{tome2010entropy}. 
Firstly, we need to find the asymptotic equilibrium state of the evolution. We consider an harmonic oscillator for simplicity, but the results are completely general. Using an approach close to the one developed in Sec.~\ref{subsec: Solution and entropy dynamics 1}, thanks to the Gaussian character of the evolution, one can describe the dynamics of such a system with a set of ordinary differential equations for the covariances of the associated quadrature operators. 
The particular structure of the effective diffusion equation in the phase space allows for the existence of an asymptotic thermal state characterized by momentum and position variance: $\sigma^2_{eq} = \frac{2D}{m \omega f}$ where $m$ and $\omega$ are the mass and frequency of the oscillator respectively and where $D$ and $f$ are the diffusion and friction constants respectively. 
Using \cref{D} we find 
\begin{equation}
\label{eq: sigmaeq}
    \sigma^2_{eq} = \frac{4}{\omega \beta}\biggr[ 1 +\frac{3\hbar^2\beta}{4mR_o^2} \biggl]\,,
\end{equation}

consequently the equilibrium state will be:

\begin{equation}\label{Weq}
    W_{eq} = \frac{1}{\pi \sigma_{eq}^2} \exp(-\frac{1}{\sigma_{eq}^2}(q^2+p^2))
\end{equation}

This result, together with the uncertainty principle that forces $\mathrm{det}(V_{eq}) \geq \hbar^2$, yields an upper bound on the parameter that determines the strength of the dissipative mechanism 
\begin{equation}
    \beta \leq \beta_c := \frac{4}{\hbar \omega - \frac{3\hbar^2}{mR^2_0}}\,.
\end{equation}
When this inequality is saturated, the asymptotic state is the ground state of the oscillator. As mentioned, $\beta$ is not the inverse energy of the noise field, but a phenomenological energy-scale of the model that determines, though, the equilibrium temperature of the dynamics. Let $\beta_{eq}$ be the inverse temperature at equilibrium, which can be found as $\sigma^2_{eq}=\dfrac{\hbar}{2}\dfrac{e^{\beta_{eq}\hbar \omega}+1}{e^{\beta_{eq}\hbar \omega}-1}$. By defining $\lambda^2_{th} = {\hbar^2 \beta}/{m}$, in the limit $\beta \ll 1$ such that $\lambda^2_{th}/R^2_0 \ll 1$, we get
\begin{equation}
\label{eq: temp}
    \dfrac{e^{\beta_{eq}\hbar \omega}+1}{e^{\beta_{eq}\hbar \omega}-1} = \frac{8}{\beta\hbar \omega} \,,
\end{equation}

which admits solution if $\frac{8}{\beta\hbar \omega} > 1$. Assuming $\beta_{eq}\hbar \omega \ll 1$ as well, we find the value $\beta_{eq} \simeq {2\beta}/({8-\beta\hbar\omega})$.

To prove that $\Pi(t)\geq 0 \,, \forall t$ we can use an equivalent expression for the entropy production rate, derived in Ref. \cite{santos2017wigner}, which is
\begin{equation} \label{eq: PI alternative}
\Pi = - \int d^3p d^3q   \calD(W_{\hat{\rho}})\ln(W_{\hat{\rho}}/W_{eq}) \,,
\end{equation}
where $\calD(W_{\hat{\rho}}) =\nabla_{\bm p}\cdot ( fpW_{\hat{\rho}})+D\Delta_{\bm p}W_{\hat{\rho}}= \nabla_{\bm p}\cdot J(W_{\hat{\rho}})$, where we introduced the current $J(W_{\hat{\rho}}) := fpW_{\hat{\rho}}+D\nabla_{\bm p}W_{\hat{\rho}}$. 

\begin{figure*}[t!]
{\bf (a)}\hskip6cm{\bf (b)}\hskip6cm{\bf (c)}\\
     \includegraphics[width=0.7\columnwidth]{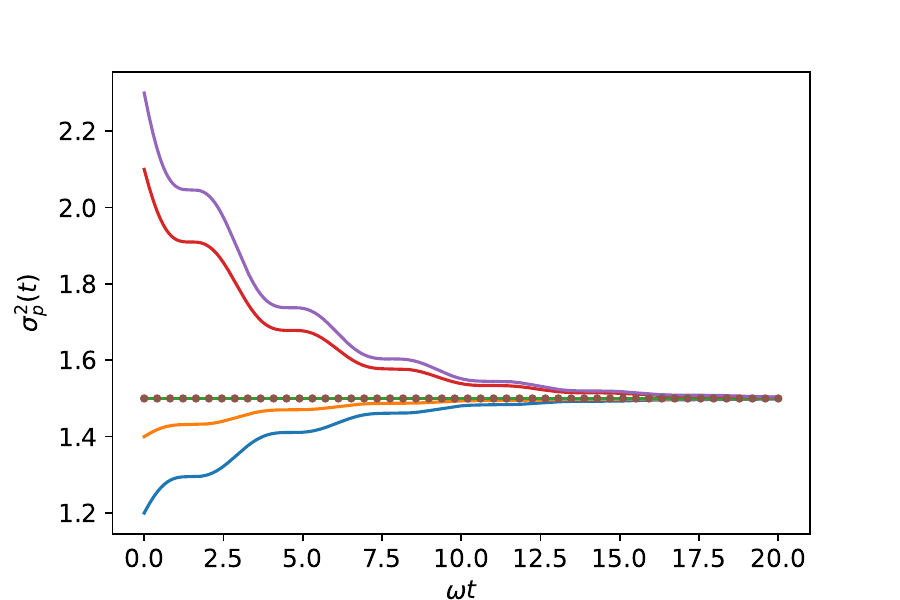}
   \includegraphics[width=0.7\columnwidth]{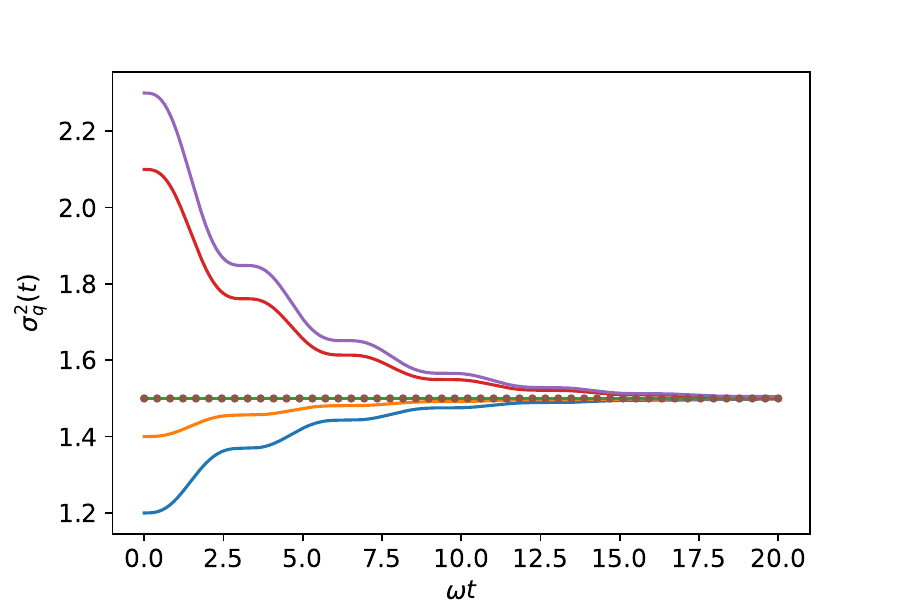}
    \includegraphics[width=0.7\columnwidth]{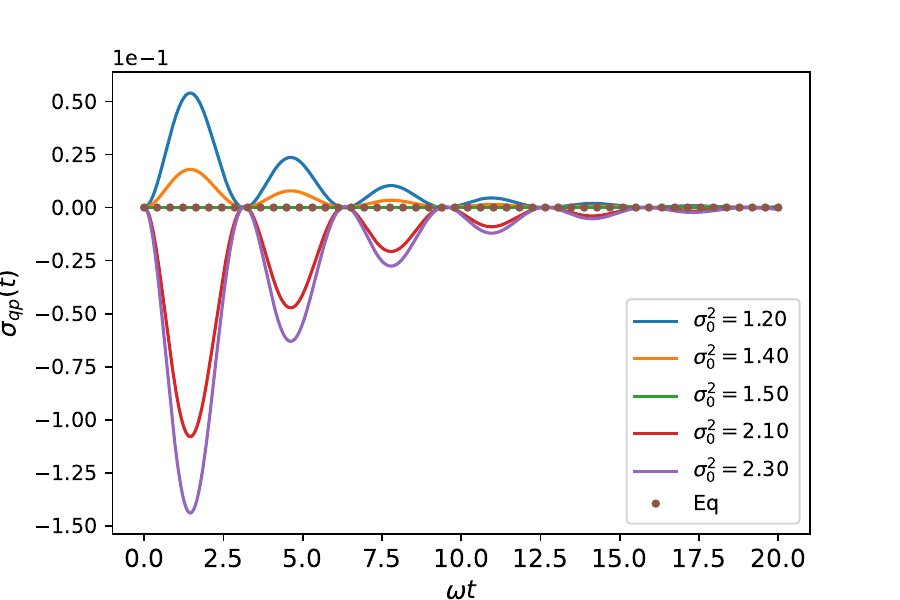}
    \caption{Evolution of the covariance matrix entries over time and their asymptotic equilibrium values.
    {\bf (a)}: Momentum variance; {\bf (b)}: Position variance {\bf (c)}: Position and Momentum covariance . Initial states at different temperatures are considered, all of them reach the equilibrium isotropic state with an associated equilibrium temperature. All parameters as in Fig.~\ref{fig:variances_fless} and we have taken $\beta = 0.3$.}
\label{fig:variances}
\end{figure*}

\begin{figure} [t!]

{\bf (a)}
    \includegraphics[width=\columnwidth]{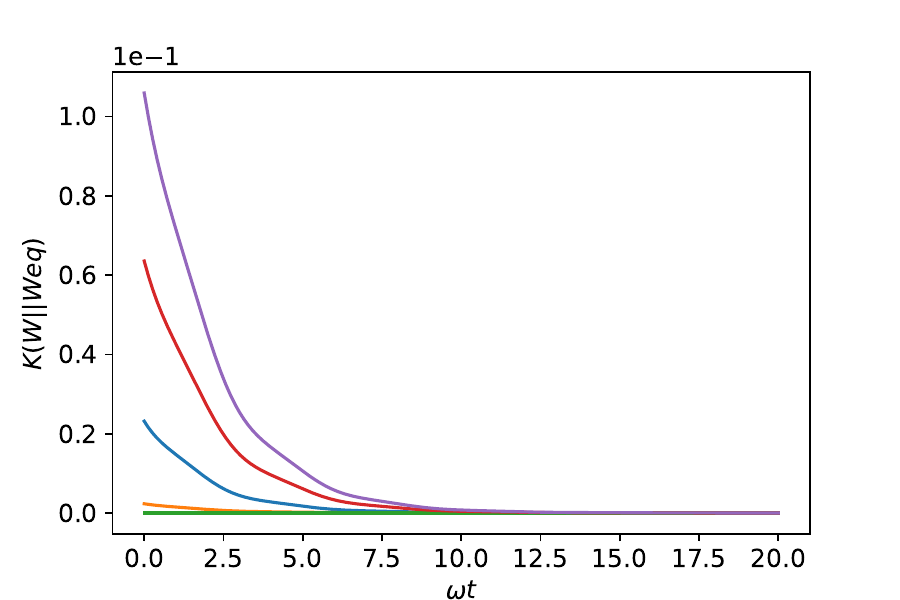}
{\bf (b)}
    \includegraphics[width=\columnwidth]{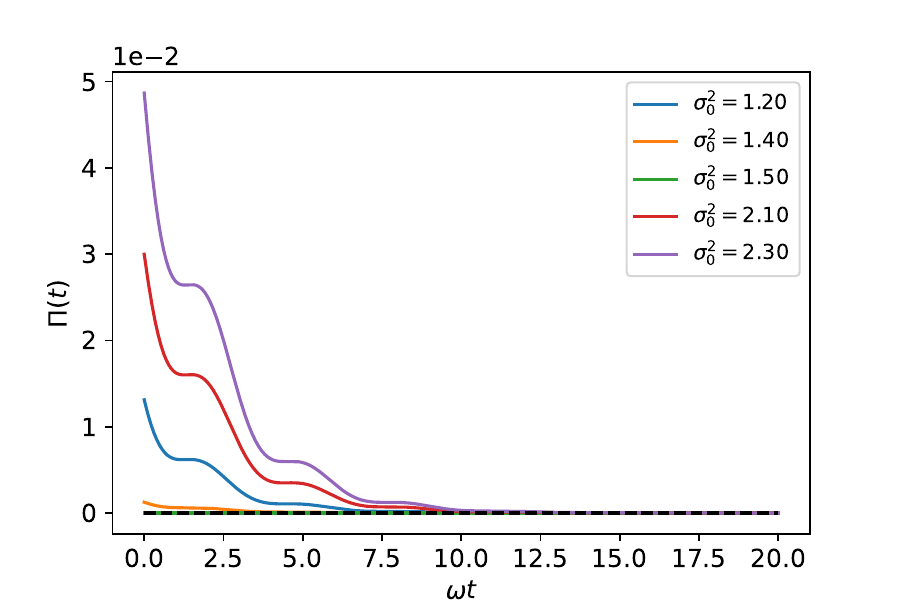}
    \caption{Relative entropy {\bf (a)} and entropy production rate {\bf (b)} varying the temperature of the initial data. In the linear friction model the system is driven towards thermal equilibrium as witnessed by the zero entropy production rate, which is otherwise always non-negative. All parameters as in Fig.~\ref{fig:variances}.}
\label{fig:entropies}
\end{figure}

Integrating by parts it is easy to see that
\begin{equation} \label{eq: nonneg PI}
\begin{aligned} 
     \Pi &= \int d^3p d^3q J(W_{\hat{\rho}}) \cdot  \biggl[ \frac{\nabla_{\bm p}W_{\hat{\rho}}}{W_{\hat{\rho}}} - \frac{\nabla_{\bm p}W_{eq}}{W_{eq}}\biggr] = \\
    & = \int d^3p d^3q \frac{J(W_{\hat{\rho}})^2}{DW_{\hat{\rho}}} \,,
\end{aligned}
\end{equation}
where in the last line we used the expression of the equilibrium state given in \cref{Weq}. Clearly, \cref{eq: nonneg PI} demonstrates that, as long as the system always has a positive Wigner function, the entropy production rate is non-negative for any initial state along the entire dynamics and equals zero when the system reaches equilibrium since $J(W_{eq})=0$. In Fig.~\ref{fig:variances} and Fig.~\ref{fig:entropies} we report the evolution of the covariances and both the relative entropy and entropy production rate respectively for initial thermal states, obtained by solving the system of ODEs. We set again $\hbar = G = 1$ and integrate over dimensionless time $\omega t$ with respect to the frequency of the oscillator over an interval of $T=20$ time units with $N=10^4$ time steps. For the sake of readability of the plots, we chose $m=2$, $R_0 = 3$, $\beta = 3$ so that $\sigma^2_{eq} = 1.5$.

\section{Non-gaussianity in the linear-friction model}
\label{sec: non gauss} 
 In order to assess non-Gaussian (thermo)dynamics stemming from large values of $\beta$, we switch to the Husimi $Q$-representation of a quantum state and deploy the Wehrl entropy  
 \begin{equation}
 S_Q = -\int d^2\alpha \, Q(\alpha)\log(Q(\alpha))
 \end{equation}
 as a quantifier of the irreversible entropy stemming from the dynamics~\cite{Santos2018}. As $Q(\alpha)$ is always non-negative, this figure of merit is always well defined, also for non-Gaussian states where the Wigner function might display negativity that would make $S_W$ imaginary~\cite{landi2021irreversible}. Since there is no guarantee that Eq. (\ref{full_W}) yields factorized solutions, one cannot trivially separate the motion in three dimensions. Moreover, studying the 3D problem is numerically demanding. For this reason, instead of focusing on the dissipative DP model, which is not well-defined in one dimension, we study the analogous CSL model \footnote{The dissipative DP model is not well-defined in one dimension because the $\Gamma(k)$ term, which is proportional to $1/k^2$ [cf.~\cref{eq: Gamma}], can be integrated in three dimensions, but leads to divergences in one dimension, much like the Newtonian potential.}. Although we are not directly examining the dissipative DP model, in this way we are nevertheless able to look at the thermodynamic properties of a class of model that share the same mathematical structure. 
 
The $Q$-representation of the quantum state $\hat\rho$ is defined as
\begin{equation}
    Q_{\hat \rho}(\alpha)=\frac{1}{\pi}\bra{\alpha}\hat \rho \ket{\alpha}
\end{equation}
with $\{\ket{\alpha}\}$ the basis of coherent states. Using the well-known phase-space correspondences

\begin{equation}
    \begin{split}
    &\bra{\alpha}a\mathcal{O}\ket{\alpha}\,=\,\left(\alpha+\frac{\partial}{\partial \alpha^\star}\right)Q_{\mathcal{O}}(\alpha)\,,\\
    &\bra{\alpha}\mathcal{O}a^\dagger\ket{\alpha}\,=\,\left(\alpha^\star+\frac{\partial}{\partial \alpha}\right)Q_{\mathcal{O}}(\alpha)\,,\\
    &\bra{\alpha}\mathcal{O}a\ket{\alpha}\,=\,\alpha\,Q_{\mathcal{O}}(\alpha)\,,\\
    &\bra{\alpha}a^\dagger\mathcal{O}\ket{\alpha}\,=
    \alpha^\star\,Q_{\mathcal{O}}(\alpha)\,.
    \end{split}
\end{equation}
and noting that with our rescaling of the phase-space variables we have: $\hat{p} = \sqrt{\hbar}i(\hat{a}^{\dagger}-\hat{a})$, $\hat{q} = \sqrt{\hbar}(\hat{a}^{\dagger}+\hat{a})$. Using this, we find

\begin{equation}
\begin{aligned}
    &\bra \alpha L_k\hat \rho L_k^\dagger \ket \alpha=\abs{m_{(+)}{-}\frac{\hbar\beta\tilde{k}}{2}\Gamma_\alpha}^2Q\left(\alpha-i\tilde{k}\right)\,,\\
&\bra\alpha\left\{ L^\dagger_k L_k,\hat\rho \right\}\ket\alpha=\left[m_{(-)}-\frac{\hbar\beta\tilde{k}}{2}\Gamma_\alpha\right]^2 Q\left(\alpha\right)+\text{c.c.},
\end{aligned}
\end{equation}
where we have set $\tilde{k}=\sqrt{\hbar}k$, $m_{(\pm)}\,=\,m\pm{\hbar\beta}\tilde{k}^2/4$ and $\Gamma_\alpha=\Im(\alpha){-}\frac{i}{2}\frac{\partial}{\partial\alpha^\star}$. We can now expand the $Q$-function for small values $\tilde{k}$, again assuming it to be well concentrated around the origin, keeping terms up to ${\cal O}(\tilde{k}^2)$ to find the $Q$-representation of the Lindblad operator 

\begin{equation}\label{QQQ}
    \mathcal{Q}(\alpha)\simeq \frac{1}{\hbar^2} \int \frac{dk}{2\pi} \Gamma(k)(A_0+A_1+A_2)
\end{equation}

where

\begin{itemize}
\item{\bf Zeroth order:}
\begin{equation}
\label{eq:Qzeroth}
    A_0=\hbar\beta m\tilde{k}^2 + \frac{\hbar^2\beta^2}{32}\tilde{k}^2\frac{\partial^2}{\partial^2\Re\alpha^2};
\end{equation}
\item{\bf First order:}
\begin{equation}
\label{eq:Qfirst}
\begin{split}
    A_1&=\hbar\beta m\tilde{k}^2\bigl(1+\frac{\hbar\beta}{4m}\tilde{k}^2\bigr)\Im\alpha\frac{\partial}{\partial\Im\alpha}
    \\
    &+ \frac{\hbar\beta}{4} m\tilde{k}^2\bigl(1+\frac{\hbar\beta}{4m}\tilde{k}^2\bigr)\frac{\partial^2}{\partial\Im\alpha^2};
\end{split}
\end{equation}
\item{\bf Second order:}
\begin{equation}
\label{eq:Qsecond}
\begin{split}
    & A_2=\frac{\hbar^2\beta^2}{16}\tilde{k}^6\frac{\partial^2}{\partial\Im\alpha^2}+\frac{\hbar\beta}{2}m\tilde{k}^4\frac{\partial^2}{\partial\Im\alpha^2}+ \frac{m^2\tilde{k}^2}{2}\frac{\partial^2}{\partial\Im\alpha^2}+\\
    & + \frac{\hbar^2\beta^2}{16}\tilde{k}^4\biggl[ \bigl(4\Im^2\alpha+1\bigr)\frac{\partial^2}{\partial\Im\alpha^2} + 2\Im\alpha \frac{\partial^3}{\partial\Im\alpha^3}\\
    &+\frac{1}{4}\biggl(\frac{\partial^4}{\partial\Im\alpha^4}+\frac{\partial^4}{\partial\Im\alpha^2\partial\Re\alpha^2}\biggr) \biggr].   
\end{split}
\end{equation}
\end{itemize}

As we found for the Wigner representation in Eq. (\ref{full_W}), we have a $4th$-order partial differential equation which is hard to solve even numerically and which will require heavy approximations in order to study.

\subsection{Early-times linearization and Second Law violation}

In order to characterize the time evolution in this regime, we can restrict ourselves to very early time to simplify the equation.
Consider a generic quantum channel: 
\begin{equation*}
    \partial_t \hat{\rho}(t) = \mathcal{L}_{TOT}[\hat{\rho}(t)]\,,
\end{equation*}
where $\mathcal{L}_{TOT} = \mathcal{L}_{1} + \mathcal{L}_{2}$, with $\mathcal{L}_{2}$ being a perturbation of the main dynamics given by $\mathcal{L}_{1}$. Let $\Phi_t[\hat{\rho}]$ be the superoperator that propagates the dynamics i.e. $\Phi_t[\hat{\rho}(0)] = \hat{\rho}(t)$ and hence it satisfies: 
\begin{equation}\label{appBPHI}
\partial_t\Phi_t = (\mathcal{L}_{1} + \mathcal{L}_{2})\Phi_t \;\;\;\; \text{with} \;\;\;\; \Phi_{t=0} = \mathbb{Id}\,.
\end{equation}
Eq.~(\ref{appBPHI}) can be cast in the integral form: 
\begin{equation}\label{formalsol}
   \Phi_t = e^{\mathcal{L}_{1}t} + \int_{0}^{t}d\tau e^{\mathcal{L}_{1}(t-\tau)}\mathcal{L}_{2}\Phi_{\tau} \,.
\end{equation}
By taking the formal solution in Eq. (\ref{formalsol}) and plugging it iteratively inside the integral one get a power series in $\mathcal{L}_{2}$. Since $\mathcal{L}_{2}$ is a perturbation, one can just keep the first order term and approximate the above expression with: 
\begin{equation}
   \Phi_t  \cong e^{\mathcal{L}_{1}t} + \int_{0}^{t}d\tau e^{\mathcal{L}_{1}(t-\tau)}\mathcal{L}_{2}e^{\mathcal{L}_{1}\tau} \,,
\end{equation}
which can be further simplified for small times $t$ (small with respect to the dynamics generated by $\mathcal{L}_{1}$) as:
\begin{equation}
\label{eq: lin_channel}
    \Phi_t  \cong e^{\mathcal{L}_{1}t} + \mathcal{L}_{2}t \,.
\end{equation}
This is finally an expression we can use to study our dynamics numerically by applying it to the QFP equation for the $Q-$function, at least at very early times. In our case $\mathcal{L}_{1}$ represents the unitary  dynamics while $\mathcal{L}_{2}$ the one related to the collapse. 
\\
In particular, assuming an initial thermal state $Q_{Th}(\alpha,\alpha^{\star}) = \dfrac{1}{\pi(1+\Bar{n})}e^{-\frac{|\alpha|^2}{1+\Bar{n}}}$, which is a fixed point of the unitary dynamics, one is left with

\begin{equation}
\begin{aligned}
     Q(t) &\cong Q_{Th}(\alpha) + \mathcal{Q}(\alpha)[Q_{Th}(\alpha)]\cdot t \; , \;\; t\ll1 \,.
\end{aligned}
\end{equation}

with $\mathcal{Q}(\alpha)$ as derived in the previous section in Eq. (\ref{QQQ}).

\begin{figure*}[t!]
\centering
\begin{minipage}[b]{0.3\textwidth}
    \centering
    \textbf{(a)}\\
    \includegraphics[width=\textwidth]{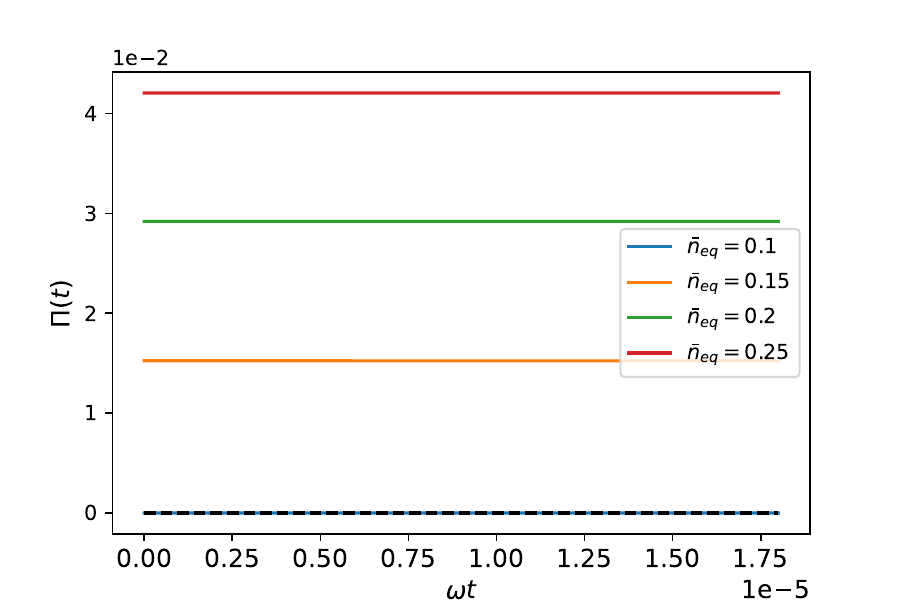}
    \vspace{-5pt}
    $\beta = 0.01$
\end{minipage}
\begin{minipage}[b]{0.3\textwidth}
    \centering
    \textbf{(b)}\\
    \includegraphics[width=\textwidth]{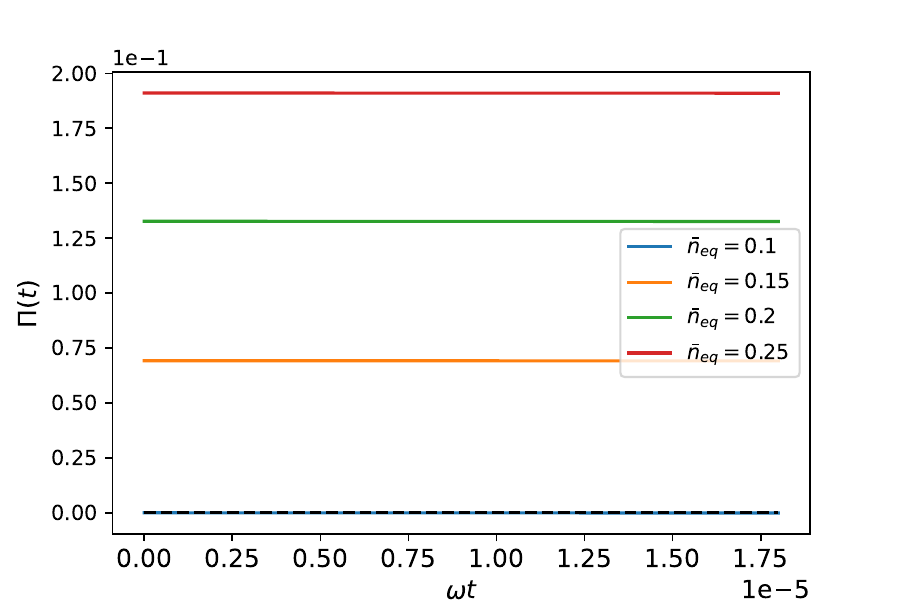}
    \vspace{-5pt}
    $\beta = 1$
\end{minipage}
\begin{minipage}[b]{0.3\textwidth}
    \centering
    \textbf{(c)}\\
    \includegraphics[width=\textwidth]{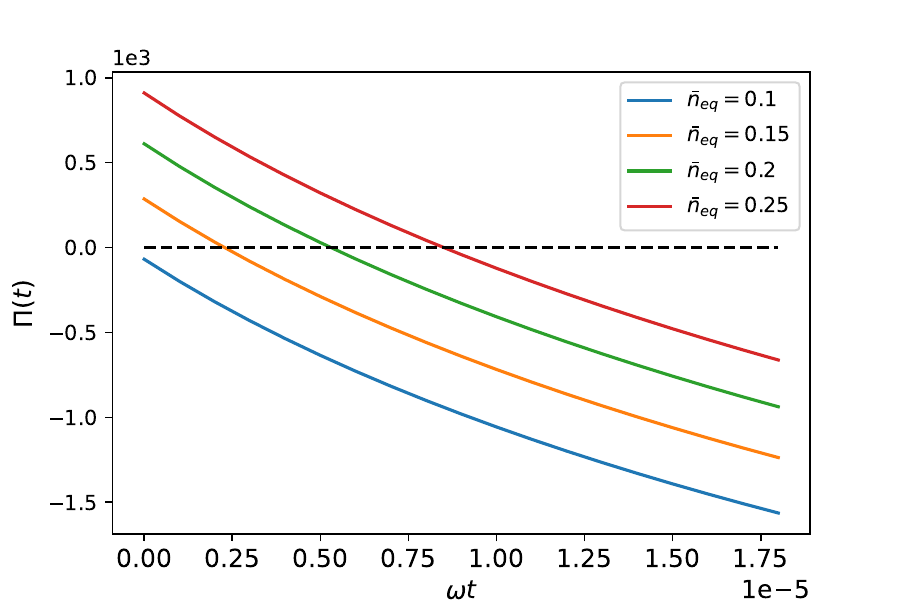}
    \vspace{-5pt}
    $\beta = 100$
\end{minipage}

\caption{Wehrl entropy production rate over time for the linearized dynamics at very early times keeping all the terms in $\beta$.
    \textbf{(a)} $\beta = 0.01$; \textbf{(b)} $\beta=1$; \textbf{(c)} $\beta=100$. Thermal initial state with $\bar{n}=0.1$ and different test target temperatures. $R_0=0.9$ and $\gamma=1$ are kept fixed as well as the mass of the oscillator $m=1$. Time is rescaled with the frequency of the oscillator. As the terms in $\beta^2$ become dominant, negative entropy production rate is witnessed.}
\label{fig:lin_pis}
\end{figure*}

As far as the entropic analysis is concerned, we must remark that we do not know if there is an asymptotic stationary state for the full Fokker-Planck equation. Anyway, following the analysis at the master equation level of Ref.~\cite{di2023linear}, we will assume that the full equation will lead to a thermalization of the state for long times and we will thus use thermal states at different temperatures to compute the entropy production rate. We have set $R_0=0.9$, $\gamma=1$ and we repeated the analysis with increasing values of $\beta$ in order to assess the effect of the non-gaussian terms on the entropy dynamics. This time we need to solve the dynamics on a grid, over which the numerical integration for the relative entropy is performed as well. We used a spatial grid on a regular lattice of $800\times 800$ points, ranging from values of $\Re\alpha$ and $\Im\alpha$ in the interval $[-4,\,4]$. Finally, we considered $40$ time steps of width $10^{-6}$ in units of the frequency of the oscillator, set $\bar{n} = 0.1$ as initial data and $\hbar{n}_{eq} = [0.1,\, 0.15,\, 0.2,\, 0.25]$ as asymptotic values since we observed a dominant heating in the early times. As always $\hbar = 1$ and $m=1$ for readability of the plots. The results are reported in Figure~\ref{fig:lin_pis}. One clearly sees that, as the terms with $\beta^2$ become dominant, the entropy production rate reaches negative values at some point, violating the Second Law of Thermodynamics. On the contrary, when they are small compared to the other terms, the dynamics remain physically consistent.
\\
This analysis suggests that large values of $\beta$ could lead to dynamics that are not physically consistent from the thermodynamics point of view.
However, the all analysis relies on the accuracy of the linearization introduced in Eq. (\ref{eq: lin_channel}), of the  which cannot be assessed the regime of validity. One would expect that for very small times Eq. (\ref{eq: lin_channel}) should be accurate, but (surprisingly) this is not always true, as discussed in details in Appendix \ref{app:linear}. Not having control on the approximation, the results in Fig. \ref{fig:lin_pis} should not be taken as set in stone, but more as an hint and motivation for further investigation in this regime. It also shows that much care as to be put when resorting to perturbation techniques; this will be the subject of future work.

\section{Conclusions}
\label{sec: concl}
We studied the dynamics of quasi-probabilities of a system subject to the gravity-related collapse mechanism described by the DP model and its linear-friction version proposed in Ref.~\cite{di2023linear}. We used this framework to compute the entropy production rate over time of the dynamics to assess the physical consistency of the models from the thermodynamic point of view. In the standard, frictionless DP model an infinite-temperature asymptotic state is required to ensure positive entropy production, causing the system to stay out-of-equilibrium indefinitely. Introducing the linear-friction term, which depends on a characteristic energy $1/\beta$, yields a Fokker-Planck equation with fourth-order derivatives that breaks the Gaussianity of the evolution, and renders a numerical solution unfeasible. In the small $\beta$ limit, Gaussianity is recovered and the resulting entropy dynamics shows a non-negative entropy production that goes to zero asymptotically, indicating a proper thermalization. In this case, we were also able to link the parameter $\beta$ to the actual asymptotic equilibrium temperature and to find an upper bound to its possible values. These findings are consistent with Ref.~\cite{artini2023characterizing} and expand upon the results therein. By retaining all terms in $\beta$, the linearization of the dynamics leads to negative entropy production, at least during the very early stages of the evolution, thereby supporting the validity of the high temperature limit. However, as discussed in Appendix \ref{app:linear}, assessing the range of validity of the linear approximation is crucial to identify the parameter range where the large-$\beta$ model is thermodynamically consistent. Nonetheless, our study paves the way towards a systematic study of collapse models outside the Gaussian-state conditions. Such investigation of the aforementioned aspects is left to future works.

\acknowledgments
We acknowledge support from
the European Union’s Horizon Europe EIC-Pathfinder
project QuCoM (101046973), the Royal Society Wolfson Fellowship (RSWF/R3/183013), the UK EPSRC (grants EP/T028424/1 and EP/X021505/1), the Department for the Economy of Northern Ireland under the US-Ireland R\&D Partnership Programme, the ``Italian National Quantum Science and Technology Institute (NQSTI)" (PE0000023) - SPOKE 2 through project ASpEQCt, the “National Centre for HPC, Big Data and Quantum Computing (HPC)” (CN00000013) – SPOKE 10 through project HyQELM, and the EU Horizon Europe EIC Pathfinder project QuCoM (GA no.~10032223), the Italian Ministry of University and Research under PNRR - M4C2-I1.3 Project PE-00000019 "HEAL ITALIA" (CUP B73C22001250006). SD acknowledges support from Istituto Nazinale di Fisica Nucleare (INFN).

\appendix

\section{Wigner function in momentum space}
\label{app:wigner_momentum}
While the standard definition of the Wigner function is the one given in Eq. \eqref{eq:wigner_position}, we derive here the alternative expression of Eq.(\ref{eq:wigner_momentum}). In order to do so, we start from \eqref{eq:wigner_position} and we insert two completeness relations:
\begin{align}
    \small W_{\hat A}(q,p)&=\int\frac{d^3y\,d^3p'd^3p''}{(4\pi)^3}e^{-i\frac{p\cdot y}{2\hbar}}\bra{q+\frac{y}{2}}\ket{p'}\times
    \\    &\times\bra{p'}\hat A \ket{p''}\bra{p''}\ket{q-\frac{y}{2}}\,\nonumber.
\end{align}
In our conventions~\cite{navarrete2022introduction} the momentum eigenstate wave fuction is normalized as $\bra{x}\ket{p}=\frac{1}{(4\pi\hbar)^{3/2}}e^{i\frac{p\cdot y}{2\hbar}}$. The expression for the Wigner function thus reduces to
\begin{align}
    \small W_{\hat A}(q,p)&=\int \frac{d^3p'd^3p''}{(4\pi\hbar)^3}\frac{d^3y}{(4\pi)^3}e^{\frac{-i y}{2\hbar}\cdot(p-p'/2-p''/2)}\times\nonumber\\
    &\times e^{\frac{i q}{2\hbar}\cdot(p'-p'')}\bra{p'}\hat A\ket{p''}\,.
\end{align}
The integral over y can be now performed explicitly, using $\int d^3y e^{i\alpha p\cdot y}=\frac{(2\pi)^3}{\abs{\alpha}^3}\delta(p)$. In our case, this generates a factor $\delta(p-p'/2-p''/2)$ and we can also perform an integration over one of the momentum integration variables, let us say $p'$:
\begin{equation}
    W_{\hat A}(q,p)=\int \frac{d^3p''}{(2\pi)^3}e^{\frac{i q}{\hbar}\cdot(p-p'')}\bra{2p-p''}\hat A \ket{p''}
\end{equation}
Performing the change of variable $p''= \frac{v}{2}+p$, we arrive to our final expression:
\begin{equation}
\label{eq:wigner_momentum_supp}
    W_{\hat A}(q,p)=\int\frac{d^3 v}{(4\pi)^3}e^{-i\frac{q\cdot v}{2\hbar}}\bra{p-\frac{v}{2}}\hat A \ket{p+\frac{v}{2}}\,,
\end{equation}
which is Eq. (\ref{eq:wigner_momentum}) of the main text.

\section{On the early time linearization}
\label{app:linear}
\begin{figure*}[t!]

\centering
\begin{minipage}[b]{0.49\textwidth}

    \centering
    \textbf{(a)}\\
    \includegraphics[width=\textwidth]{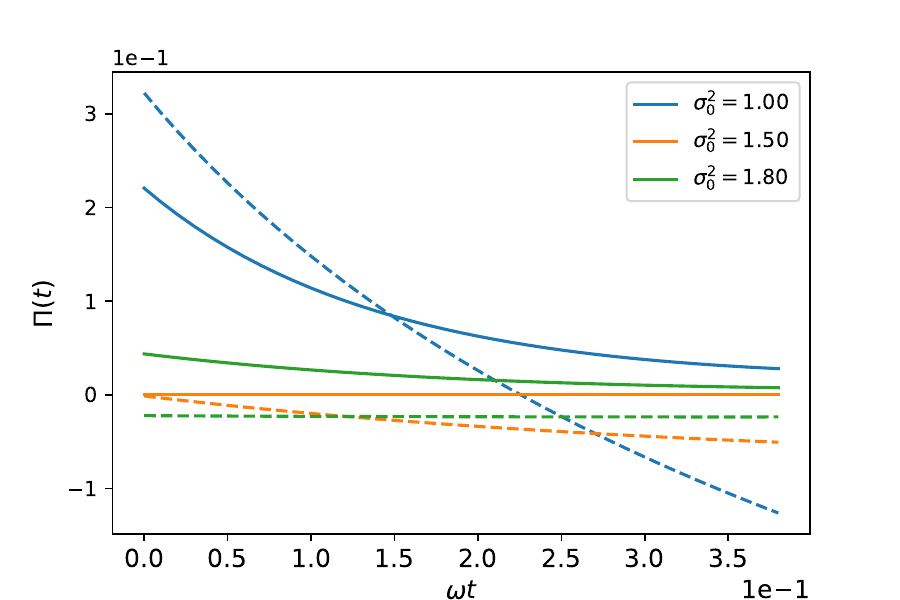}
    \vspace{-5pt}
    $D=1$, $f=4/3$
    
\end{minipage}
\begin{minipage}[b]{0.49\textwidth}
    \centering
    \textbf{(b)}\\
    \includegraphics[width=\textwidth]{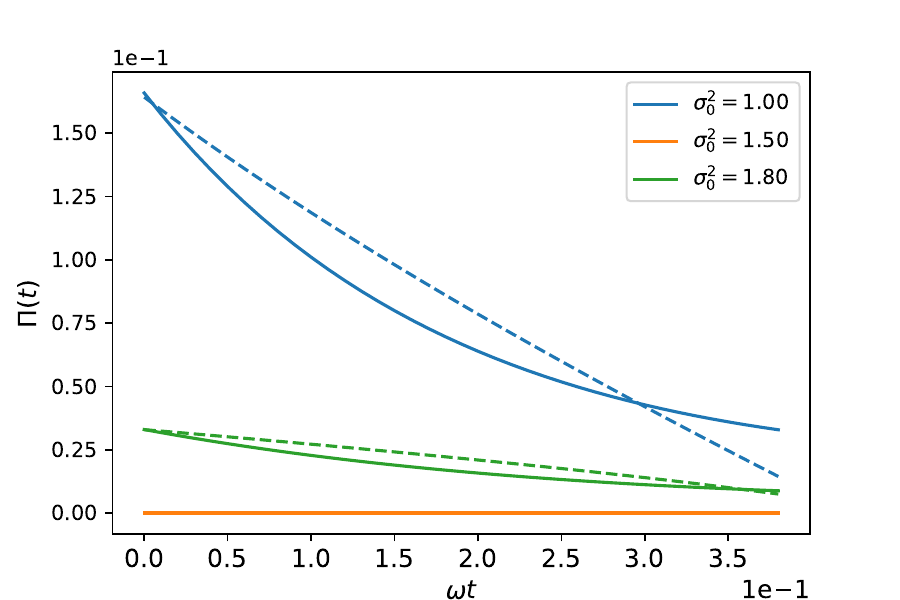}
    \vspace{-5pt}
    $D=3/4$, $f=1$
\end{minipage}
\caption{Comparison between the entropy production rate computed with the exact solution (solid lines) and the one computed with the linearized solution (dashed lines) for different initial states. The parameters are chosen so that the asymptotic state is the same. 
    {\bf (a)}: The diffusion and dissipation parameters are $D=1$, $f=4/3$. There is a discrepancy between the two solutions from $t=0$, initial negative entropy production is even present. The linearization fails; {\bf (b)}: The diffusion and dissipation parameters are $D=3/4$, $f=1$ for very early times the two methods are in excellent agreement.}
\label{fig: linear_comp}
\end{figure*}

\begin{figure*}[t!]

\centering
\begin{minipage}[b]{0.49\textwidth}
    \centering
    \textbf{(a)}\\
    \includegraphics[width=\textwidth]{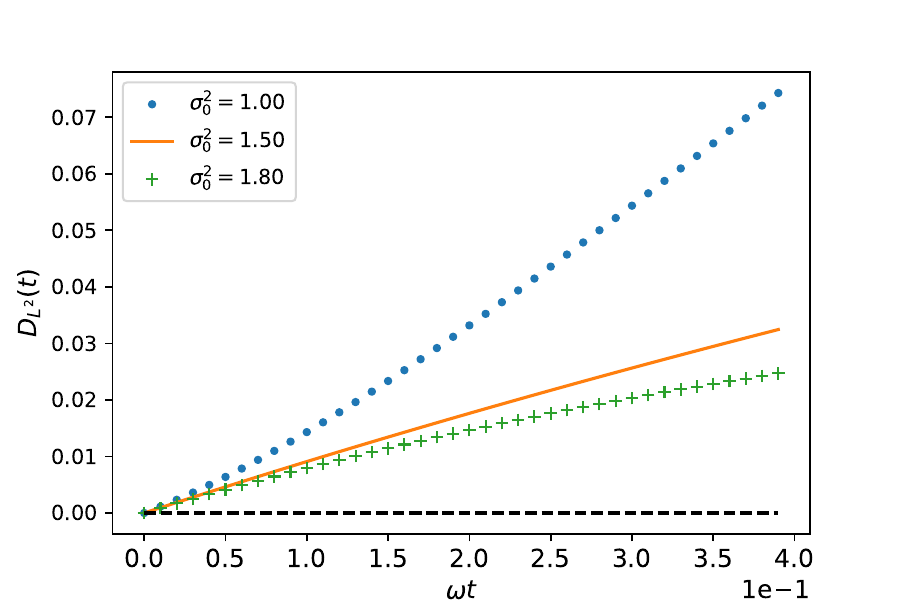}
    \vspace{-5pt}
    $D=1$, $f=4/3$
\end{minipage}
\begin{minipage}[b]{0.49\textwidth}
    \centering
    \textbf{(b)}\\
    \includegraphics[width=\textwidth]{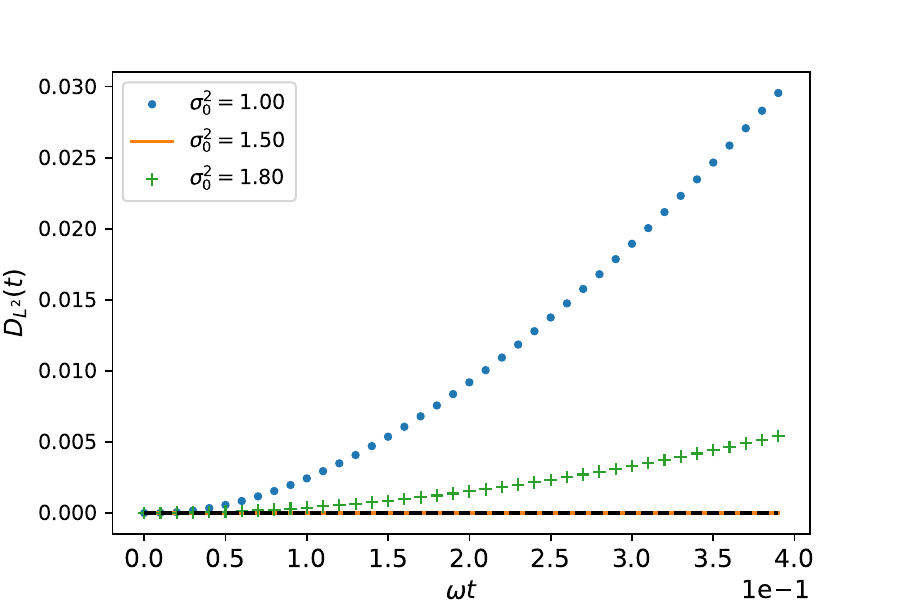}
    \vspace{-5pt}
    $D=3/4$, $f=1$
\end{minipage}
\caption{ $L_2$-norm distance between the exact and the linearized state, computed as $D_{L_2} = \sqrt{\int dqdp(W_{lin}(q,p)-W_{ex}(q,p))^2}$
For $D=1$, $f=4/3$ the approximate Wigner diverges from the exact one much faster than for $D=3/4$, $f=1$. In particular, the stationary state is not stationary for the linearized dynamic with $D=1$, $f=4/3$.}
\label{fig: L2norm}
\end{figure*}

In this Appendix we elaborate on the limitation of the early times expansion performed in \cref{sec: non gauss}. In order to do so we will perform the same linearization to \cref{eq: smallBetaQFP} and compare it with the exact solution. In this section we will consider only the one-dimensional case and we will treat the diffusion coefficient, $D$, and the friction coefficient, $f$, as free parameters since we are only interested in the features of the linearization procedure. 
\\
We consider again the linearized version of the full quantum channel as in \cref{eq: lin_channel} where $\calL_1$ is still the unitary part, whereas $\calL_2$ is the noise, perturbative term, in the small $\beta$ limit. Considering an initial thermal state, then, the linearized evolution of the Wigner function of the system yields:
\begin{equation}
    W(q,p,t) \cong W_{Th}(q,p) + \mathcal{W}(q,p)[W_{Th}(q,p)]\cdot t \,,
\end{equation}
with 
\begin{equation}
    \mathcal{W}(q,p)[W] = \partial_p(fpW) + D\Delta_PW \,.
\end{equation}
We compute the approximate evolution on a regular lattice of $800\times 800$ points where both $p$ and $q$ range in the interval $[-4,4]$ and considering a time interval $T=40$ with time steps $dt=10^{-2}$, as always rescaled with the frequency of the oscillator. In Fig. \ref{fig: linear_comp} we compare the results of the linearized evolution for the entropy production rate with those of the exact numerical solution. One is able to see that for certain choices of parameters the linearization is indeed able to capture the early times feaures of the entropy dynamics, but this is not always true as seen in Fig. \ref{fig: linear_comp}$\textbf{(a)}$, where the linearized solution gives different results and even negative entropy production rate. These different choices of parameters thus give different reliability of the linearization method, as it is also confirmed by Fig. \ref{fig: L2norm} where the $L_2$-norm distance between the exact Wigner function at time $t$ and the linearized one is reported.
\\
To summarize, this analysis shows that the linearization of the Lindbladian in Eq. (\ref{eq: lin_channel}) is not always a reliable tool to approximate the early time evolutions of quantum systems, as it could lead to sensible differences with the exact solutions upon certain choices of the parameters of the dynamics. Furthermore, we certified that the Wigner functions generated by the linearized evolution are indeed proper quantum states (i.e. their covariance matrix satisfy the Heisenberg uncertainty principle), hence the negative entropy production rate must be a feature of such an approximate dynamics that signals its imprecision. Nevertheless, since in some regimes the approximate dynamics is accurate, it might still be that for large values of $\beta$ the model leads to negative entropy production. Having no control on the approximation, we cannot conclude anything. The high sensitivity of the approximation in Eq. (\ref{eq: lin_channel}) from the the choice of parameters and the fact that the approximation, in some cases, fails also for ly small times is an intriguing feature, which we believe may be of interest for the readers. Understanding rigorously the range of validity of this approximation is not a trivial task, which we leave for future study.

\newpage
\bibliography{bibliography}

\begin{thebibliography}{45}%
\makeatletter
\providecommand \@ifxundefined [1]{%
 \@ifx{#1\undefined}
}%
\providecommand \@ifnum [1]{%
 \ifnum #1\expandafter \@firstoftwo
 \else \expandafter \@secondoftwo
 \fi
}%
\providecommand \@ifx [1]{%
 \ifx #1\expandafter \@firstoftwo
 \else \expandafter \@secondoftwo
 \fi
}%
\providecommand \natexlab [1]{#1}%
\providecommand \enquote  [1]{``#1''}%
\providecommand \bibnamefont  [1]{#1}%
\providecommand \bibfnamefont [1]{#1}%
\providecommand \citenamefont [1]{#1}%
\providecommand \href@noop [0]{\@secondoftwo}%
\providecommand \href [0]{\begingroup \@sanitize@url \@href}%
\providecommand \@href[1]{\@@startlink{#1}\@@href}%
\providecommand \@@href[1]{\endgroup#1\@@endlink}%
\providecommand \@sanitize@url [0]{\catcode `\\12\catcode `\$12\catcode `\&12\catcode `\#12\catcode `\^12\catcode `\_12\catcode `\%12\relax}%
\providecommand \@@startlink[1]{}%
\providecommand \@@endlink[0]{}%
\providecommand \url  [0]{\begingroup\@sanitize@url \@url }%
\providecommand \@url [1]{\endgroup\@href {#1}{\urlprefix }}%
\providecommand \urlprefix  [0]{URL }%
\providecommand \Eprint [0]{\href }%
\providecommand \doibase [0]{https://doi.org/}%
\providecommand \selectlanguage [0]{\@gobble}%
\providecommand \bibinfo  [0]{\@secondoftwo}%
\providecommand \bibfield  [0]{\@secondoftwo}%
\providecommand \translation [1]{[#1]}%
\providecommand \BibitemOpen [0]{}%
\providecommand \bibitemStop [0]{}%
\providecommand \bibitemNoStop [0]{.\EOS\space}%
\providecommand \EOS [0]{\spacefactor3000\relax}%
\providecommand \BibitemShut  [1]{\csname bibitem#1\endcsname}%
\let\auto@bib@innerbib\@empty
\bibitem [{\citenamefont {Penrose}(1996)}]{penrose1996gravity}%
  \BibitemOpen
  \bibfield  {author} {\bibinfo {author} {\bibfnamefont {R.}~\bibnamefont {Penrose}},\ }\bibfield  {title} {\bibinfo {title} {On gravity's role in quantum state reduction},\ }\href {https://doi.org/10.1007/BF02105068} {\bibfield  {journal} {\bibinfo  {journal} {General relativity and gravitation}\ }\textbf {\bibinfo {volume} {28}},\ \bibinfo {pages} {581} (\bibinfo {year} {1996})}\BibitemShut {NoStop}%
\bibitem [{\citenamefont {Bassi}\ \emph {et~al.}(2013)\citenamefont {Bassi}, \citenamefont {Lochan}, \citenamefont {Satin}, \citenamefont {Singh},\ and\ \citenamefont {Ulbricht}}]{bassi2013models}%
  \BibitemOpen
  \bibfield  {author} {\bibinfo {author} {\bibfnamefont {A.}~\bibnamefont {Bassi}}, \bibinfo {author} {\bibfnamefont {K.}~\bibnamefont {Lochan}}, \bibinfo {author} {\bibfnamefont {S.}~\bibnamefont {Satin}}, \bibinfo {author} {\bibfnamefont {T.~P.}\ \bibnamefont {Singh}},\ and\ \bibinfo {author} {\bibfnamefont {H.}~\bibnamefont {Ulbricht}},\ }\bibfield  {title} {\bibinfo {title} {Models of wave-function collapse, underlying theories, and experimental tests},\ }\href {https://doi.org/10.1103/RevModPhys.85.471} {\bibfield  {journal} {\bibinfo  {journal} {Reviews of Modern Physics}\ }\textbf {\bibinfo {volume} {85}},\ \bibinfo {pages} {471} (\bibinfo {year} {2013})}\BibitemShut {NoStop}%
\bibitem [{\citenamefont {Karolyhazy}(1966)}]{karolyhazy1966gravitation}%
  \BibitemOpen
  \bibfield  {author} {\bibinfo {author} {\bibfnamefont {F.}~\bibnamefont {Karolyhazy}},\ }\bibfield  {title} {\bibinfo {title} {Gravitation and quantum mechanics of macroscopic objects},\ }\href {https://doi.org/10.1007/BF02717926} {\bibfield  {journal} {\bibinfo  {journal} {Il Nuovo Cimento A (1965-1970)}\ }\textbf {\bibinfo {volume} {42}},\ \bibinfo {pages} {390} (\bibinfo {year} {1966})}\BibitemShut {NoStop}%
\bibitem [{\citenamefont {Diosi}(1987)}]{diosi1987universal}%
  \BibitemOpen
  \bibfield  {author} {\bibinfo {author} {\bibfnamefont {L.}~\bibnamefont {Diosi}},\ }\bibfield  {title} {\bibinfo {title} {A universal master equation for the gravitational violation of quantum mechanics},\ }\href {https://doi.org/10.1016/0375-9601(87)90681-5} {\bibfield  {journal} {\bibinfo  {journal} {Physics Letters A}\ }\textbf {\bibinfo {volume} {120}},\ \bibinfo {pages} {377} (\bibinfo {year} {1987})}\BibitemShut {NoStop}%
\bibitem [{\citenamefont {Di{\'o}si}(1989)}]{diosi1989models}%
  \BibitemOpen
  \bibfield  {author} {\bibinfo {author} {\bibfnamefont {L.}~\bibnamefont {Di{\'o}si}},\ }\bibfield  {title} {\bibinfo {title} {Models for universal reduction of macroscopic quantum fluctuations},\ }\href {https://doi.org/10.1103/PhysRevA.40.1165} {\bibfield  {journal} {\bibinfo  {journal} {Physical Review A}\ }\textbf {\bibinfo {volume} {40}},\ \bibinfo {pages} {1165} (\bibinfo {year} {1989})}\BibitemShut {NoStop}%
\bibitem [{\citenamefont {Hu}\ and\ \citenamefont {Verdaguer}(2008)}]{hu2008stochastic}%
  \BibitemOpen
  \bibfield  {author} {\bibinfo {author} {\bibfnamefont {B.~L.}\ \bibnamefont {Hu}}\ and\ \bibinfo {author} {\bibfnamefont {E.}~\bibnamefont {Verdaguer}},\ }\bibfield  {title} {\bibinfo {title} {Stochastic gravity: Theory and applications},\ }\href {https://doi.org/10.12942/lrr-2008-3} {\bibfield  {journal} {\bibinfo  {journal} {Living Reviews in Relativity}\ }\textbf {\bibinfo {volume} {11}},\ \bibinfo {pages} {1} (\bibinfo {year} {2008})}\BibitemShut {NoStop}%
\bibitem [{\citenamefont {Breuer}\ \emph {et~al.}(2009)\citenamefont {Breuer}, \citenamefont {G{\"o}kl{\"u}},\ and\ \citenamefont {L{\"a}mmerzahl}}]{breuer2009metric}%
  \BibitemOpen
  \bibfield  {author} {\bibinfo {author} {\bibfnamefont {H.-P.}\ \bibnamefont {Breuer}}, \bibinfo {author} {\bibfnamefont {E.}~\bibnamefont {G{\"o}kl{\"u}}},\ and\ \bibinfo {author} {\bibfnamefont {C.}~\bibnamefont {L{\"a}mmerzahl}},\ }\bibfield  {title} {\bibinfo {title} {Metric fluctuations and decoherence},\ }\href {https://doi.org/10.1088/0264-9381/26/10/105012} {\bibfield  {journal} {\bibinfo  {journal} {Classical and Quantum Gravity}\ }\textbf {\bibinfo {volume} {26}},\ \bibinfo {pages} {105012} (\bibinfo {year} {2009})}\BibitemShut {NoStop}%
\bibitem [{\citenamefont {Blencowe}(2013)}]{blencowe2013effective}%
  \BibitemOpen
  \bibfield  {author} {\bibinfo {author} {\bibfnamefont {M.}~\bibnamefont {Blencowe}},\ }\bibfield  {title} {\bibinfo {title} {Effective field theory approach to gravitationally induced decoherence},\ }\href {https://doi.org/10.1103/PhysRevLett.111.021302} {\bibfield  {journal} {\bibinfo  {journal} {Physical Review Letters}\ }\textbf {\bibinfo {volume} {111}},\ \bibinfo {pages} {021302} (\bibinfo {year} {2013})}\BibitemShut {NoStop}%
\bibitem [{\citenamefont {Gasbarri}\ \emph {et~al.}(2017)\citenamefont {Gasbarri}, \citenamefont {Toro{\v{s}}}, \citenamefont {Donadi},\ and\ \citenamefont {Bassi}}]{gasbarri2017gravity}%
  \BibitemOpen
  \bibfield  {author} {\bibinfo {author} {\bibfnamefont {G.}~\bibnamefont {Gasbarri}}, \bibinfo {author} {\bibfnamefont {M.}~\bibnamefont {Toro{\v{s}}}}, \bibinfo {author} {\bibfnamefont {S.}~\bibnamefont {Donadi}},\ and\ \bibinfo {author} {\bibfnamefont {A.}~\bibnamefont {Bassi}},\ }\bibfield  {title} {\bibinfo {title} {Gravity induced wave function collapse},\ }\href {https://doi.org/10.1103/PhysRevD.96.104013} {\bibfield  {journal} {\bibinfo  {journal} {Physical Review D}\ }\textbf {\bibinfo {volume} {96}},\ \bibinfo {pages} {104013} (\bibinfo {year} {2017})}\BibitemShut {NoStop}%
\bibitem [{\citenamefont {Asprea}\ \emph {et~al.}(2021)\citenamefont {Asprea}, \citenamefont {Gasbarri},\ and\ \citenamefont {Bassi}}]{asprea2021gravitational}%
  \BibitemOpen
  \bibfield  {author} {\bibinfo {author} {\bibfnamefont {L.}~\bibnamefont {Asprea}}, \bibinfo {author} {\bibfnamefont {G.}~\bibnamefont {Gasbarri}},\ and\ \bibinfo {author} {\bibfnamefont {A.}~\bibnamefont {Bassi}},\ }\bibfield  {title} {\bibinfo {title} {Gravitational decoherence: A general nonrelativistic model},\ }\href {https://doi.org/10.1103/PhysRevD.103.104041} {\bibfield  {journal} {\bibinfo  {journal} {Physical Review D}\ }\textbf {\bibinfo {volume} {103}},\ \bibinfo {pages} {104041} (\bibinfo {year} {2021})}\BibitemShut {NoStop}%
\bibitem [{\citenamefont {Bassi}\ \emph {et~al.}(2017)\citenamefont {Bassi}, \citenamefont {Gro{\ss}ardt},\ and\ \citenamefont {Ulbricht}}]{bassi2017gravitational}%
  \BibitemOpen
  \bibfield  {author} {\bibinfo {author} {\bibfnamefont {A.}~\bibnamefont {Bassi}}, \bibinfo {author} {\bibfnamefont {A.}~\bibnamefont {Gro{\ss}ardt}},\ and\ \bibinfo {author} {\bibfnamefont {H.}~\bibnamefont {Ulbricht}},\ }\bibfield  {title} {\bibinfo {title} {Gravitational decoherence},\ }\href {https://iopscience.iop.org/article/10.1088/1361-6382/aa864f} {\bibfield  {journal} {\bibinfo  {journal} {Classical and Quantum Gravity}\ }\textbf {\bibinfo {volume} {34}},\ \bibinfo {pages} {193002} (\bibinfo {year} {2017})}\BibitemShut {NoStop}%
\bibitem [{\citenamefont {Di{\'o}si}\ and\ \citenamefont {Halliwell}(1998)}]{diosi1998coupling}%
  \BibitemOpen
  \bibfield  {author} {\bibinfo {author} {\bibfnamefont {L.}~\bibnamefont {Di{\'o}si}}\ and\ \bibinfo {author} {\bibfnamefont {J.~J.}\ \bibnamefont {Halliwell}},\ }\bibfield  {title} {\bibinfo {title} {Coupling classical and quantum variables using continuous quantum measurement theory},\ }\href {https://doi.org/10.1103/PhysRevLett.81.2846} {\bibfield  {journal} {\bibinfo  {journal} {Physical Review Letters}\ }\textbf {\bibinfo {volume} {81}},\ \bibinfo {pages} {2846} (\bibinfo {year} {1998})}\BibitemShut {NoStop}%
\bibitem [{\citenamefont {Di{\'o}si}\ \emph {et~al.}(2000)\citenamefont {Di{\'o}si}, \citenamefont {Gisin},\ and\ \citenamefont {Strunz}}]{diosi2000quantum}%
  \BibitemOpen
  \bibfield  {author} {\bibinfo {author} {\bibfnamefont {L.}~\bibnamefont {Di{\'o}si}}, \bibinfo {author} {\bibfnamefont {N.}~\bibnamefont {Gisin}},\ and\ \bibinfo {author} {\bibfnamefont {W.~T.}\ \bibnamefont {Strunz}},\ }\bibfield  {title} {\bibinfo {title} {Quantum approach to coupling classical and quantum dynamics},\ }\href {https://doi.org/10.1103/PhysRevA.61.022108} {\bibfield  {journal} {\bibinfo  {journal} {Physical Review A}\ }\textbf {\bibinfo {volume} {61}},\ \bibinfo {pages} {022108} (\bibinfo {year} {2000})}\BibitemShut {NoStop}%
\bibitem [{\citenamefont {Di{\'o}si}(2023)}]{diosi2023hybrid}%
  \BibitemOpen
  \bibfield  {author} {\bibinfo {author} {\bibfnamefont {L.}~\bibnamefont {Di{\'o}si}},\ }\bibfield  {title} {\bibinfo {title} {Hybrid completely positive markovian quantum-classical dynamics},\ }\href {https://doi.org/10.1103/PhysRevA.107.062206} {\bibfield  {journal} {\bibinfo  {journal} {Physical Review A}\ }\textbf {\bibinfo {volume} {107}},\ \bibinfo {pages} {062206} (\bibinfo {year} {2023})}\BibitemShut {NoStop}%
\bibitem [{\citenamefont {Oppenheim}(2023)}]{oppenheim2023postquantum}%
  \BibitemOpen
  \bibfield  {author} {\bibinfo {author} {\bibfnamefont {J.}~\bibnamefont {Oppenheim}},\ }\bibfield  {title} {\bibinfo {title} {A postquantum theory of classical gravity?},\ }\href {https://doi.org/10.1103/PhysRevX.13.041040} {\bibfield  {journal} {\bibinfo  {journal} {Physical Review X}\ }\textbf {\bibinfo {volume} {13}},\ \bibinfo {pages} {041040} (\bibinfo {year} {2023})}\BibitemShut {NoStop}%
\bibitem [{\citenamefont {Donadi}\ \emph {et~al.}(2023)\citenamefont {Donadi}, \citenamefont {Ferialdi},\ and\ \citenamefont {Bassi}}]{donadi2023collapse}%
  \BibitemOpen
  \bibfield  {author} {\bibinfo {author} {\bibfnamefont {S.}~\bibnamefont {Donadi}}, \bibinfo {author} {\bibfnamefont {L.}~\bibnamefont {Ferialdi}},\ and\ \bibinfo {author} {\bibfnamefont {A.}~\bibnamefont {Bassi}},\ }\bibfield  {title} {\bibinfo {title} {Collapse dynamics are diffusive},\ }\href {https://doi.org/10.1103/PhysRevLett.130.230202} {\bibfield  {journal} {\bibinfo  {journal} {Physical Review Letters}\ }\textbf {\bibinfo {volume} {130}},\ \bibinfo {pages} {230202} (\bibinfo {year} {2023})}\BibitemShut {NoStop}%
\bibitem [{\citenamefont {Vinante}\ and\ \citenamefont {Ulbricht}(2021)}]{vinante2021gravity}%
  \BibitemOpen
  \bibfield  {author} {\bibinfo {author} {\bibfnamefont {A.}~\bibnamefont {Vinante}}\ and\ \bibinfo {author} {\bibfnamefont {H.}~\bibnamefont {Ulbricht}},\ }\bibfield  {title} {\bibinfo {title} {Gravity-related collapse of the wave function and spontaneous heating: Revisiting the experimental bounds},\ }\href {https://doi.org/10.1116/5.0073450} {\bibfield  {journal} {\bibinfo  {journal} {AVS Quantum Science}\ }\textbf {\bibinfo {volume} {3}} (\bibinfo {year} {2021})}\BibitemShut {NoStop}%
\bibitem [{\citenamefont {Donadi}\ \emph {et~al.}(2021)\citenamefont {Donadi}, \citenamefont {Piscicchia}, \citenamefont {Curceanu}, \citenamefont {Di{\'o}si}, \citenamefont {Laubenstein},\ and\ \citenamefont {Bassi}}]{donadi2021underground}%
  \BibitemOpen
  \bibfield  {author} {\bibinfo {author} {\bibfnamefont {S.}~\bibnamefont {Donadi}}, \bibinfo {author} {\bibfnamefont {K.}~\bibnamefont {Piscicchia}}, \bibinfo {author} {\bibfnamefont {C.}~\bibnamefont {Curceanu}}, \bibinfo {author} {\bibfnamefont {L.}~\bibnamefont {Di{\'o}si}}, \bibinfo {author} {\bibfnamefont {M.}~\bibnamefont {Laubenstein}},\ and\ \bibinfo {author} {\bibfnamefont {A.}~\bibnamefont {Bassi}},\ }\bibfield  {title} {\bibinfo {title} {Underground test of gravity-related wave function collapse},\ }\href {https://doi.org/10.1038/s41567-020-1008-4} {\bibfield  {journal} {\bibinfo  {journal} {Nature Physics}\ }\textbf {\bibinfo {volume} {17}},\ \bibinfo {pages} {74} (\bibinfo {year} {2021})}\BibitemShut {NoStop}%
\bibitem [{\citenamefont {Piscicchia}\ \emph {et~al.}(2024)\citenamefont {Piscicchia}, \citenamefont {Donadi}, \citenamefont {Manti}, \citenamefont {Bassi}, \citenamefont {Derakhshani}, \citenamefont {Di{\'o}si},\ and\ \citenamefont {Curceanu}}]{piscicchia2024x}%
  \BibitemOpen
  \bibfield  {author} {\bibinfo {author} {\bibfnamefont {K.}~\bibnamefont {Piscicchia}}, \bibinfo {author} {\bibfnamefont {S.}~\bibnamefont {Donadi}}, \bibinfo {author} {\bibfnamefont {S.}~\bibnamefont {Manti}}, \bibinfo {author} {\bibfnamefont {A.}~\bibnamefont {Bassi}}, \bibinfo {author} {\bibfnamefont {M.}~\bibnamefont {Derakhshani}}, \bibinfo {author} {\bibfnamefont {L.}~\bibnamefont {Di{\'o}si}},\ and\ \bibinfo {author} {\bibfnamefont {C.}~\bibnamefont {Curceanu}},\ }\bibfield  {title} {\bibinfo {title} {X-ray emission from atomic systems can distinguish between prevailing dynamical wave-function collapse models},\ }\href {https://doi.org/10.1103/PhysRevLett.132.250203} {\bibfield  {journal} {\bibinfo  {journal} {Physical Review Letters}\ }\textbf {\bibinfo {volume} {132}},\ \bibinfo {pages} {250203} (\bibinfo {year} {2024})}\BibitemShut {NoStop}%
\bibitem [{\citenamefont {Arnquist}\ \emph {et~al.}(2022)\citenamefont {Arnquist}, \citenamefont {Avignone~III}, \citenamefont {Barabash}, \citenamefont {Barton}, \citenamefont {Bhimani}, \citenamefont {Blalock}, \citenamefont {Bos}, \citenamefont {Busch}, \citenamefont {Buuck}, \citenamefont {Caldwell} \emph {et~al.}}]{arnquist2022search}%
  \BibitemOpen
  \bibfield  {author} {\bibinfo {author} {\bibfnamefont {I.}~\bibnamefont {Arnquist}}, \bibinfo {author} {\bibfnamefont {F.}~\bibnamefont {Avignone~III}}, \bibinfo {author} {\bibfnamefont {A.}~\bibnamefont {Barabash}}, \bibinfo {author} {\bibfnamefont {C.}~\bibnamefont {Barton}}, \bibinfo {author} {\bibfnamefont {K.}~\bibnamefont {Bhimani}}, \bibinfo {author} {\bibfnamefont {E.}~\bibnamefont {Blalock}}, \bibinfo {author} {\bibfnamefont {B.}~\bibnamefont {Bos}}, \bibinfo {author} {\bibfnamefont {M.}~\bibnamefont {Busch}}, \bibinfo {author} {\bibfnamefont {M.}~\bibnamefont {Buuck}}, \bibinfo {author} {\bibfnamefont {T.}~\bibnamefont {Caldwell}}, \emph {et~al.},\ }\bibfield  {title} {\bibinfo {title} {Search for spontaneous radiation from wave function collapse in the majorana demonstrator},\ }\href {https://doi.org/10.1103/PhysRevLett.129.080401} {\bibfield  {journal} {\bibinfo  {journal} {Physical Review Letters}\ }\textbf {\bibinfo {volume} {129}},\ \bibinfo {pages} {080401} (\bibinfo {year}
  {2022})}\BibitemShut {NoStop}%
\bibitem [{\citenamefont {Smirne}\ and\ \citenamefont {Bassi}(2015)}]{smirne2015dissipative}%
  \BibitemOpen
  \bibfield  {author} {\bibinfo {author} {\bibfnamefont {A.}~\bibnamefont {Smirne}}\ and\ \bibinfo {author} {\bibfnamefont {A.}~\bibnamefont {Bassi}},\ }\bibfield  {title} {\bibinfo {title} {Dissipative continuous spontaneous localization ({CSL}) model},\ }\href {https://doi.org/10.1038/srep12518} {\bibfield  {journal} {\bibinfo  {journal} {Scientific reports}\ }\textbf {\bibinfo {volume} {5}},\ \bibinfo {pages} {12518} (\bibinfo {year} {2015})}\BibitemShut {NoStop}%
\bibitem [{\citenamefont {Di~Bartolomeo}\ \emph {et~al.}(2023)\citenamefont {Di~Bartolomeo}, \citenamefont {Carlesso}, \citenamefont {Piscicchia}, \citenamefont {Curceanu}, \citenamefont {Derakhshani},\ and\ \citenamefont {Di\'osi}}]{di2023linear}%
  \BibitemOpen
  \bibfield  {author} {\bibinfo {author} {\bibfnamefont {G.}~\bibnamefont {Di~Bartolomeo}}, \bibinfo {author} {\bibfnamefont {M.}~\bibnamefont {Carlesso}}, \bibinfo {author} {\bibfnamefont {K.}~\bibnamefont {Piscicchia}}, \bibinfo {author} {\bibfnamefont {C.}~\bibnamefont {Curceanu}}, \bibinfo {author} {\bibfnamefont {M.}~\bibnamefont {Derakhshani}},\ and\ \bibinfo {author} {\bibfnamefont {L.}~\bibnamefont {Di\'osi}},\ }\bibfield  {title} {\bibinfo {title} {Linear-friction many-body equation for dissipative spontaneous wave-function collapse},\ }\href {https://doi.org/10.1103/PhysRevA.108.012202} {\bibfield  {journal} {\bibinfo  {journal} {Phys. Rev. A}\ }\textbf {\bibinfo {volume} {108}},\ \bibinfo {pages} {012202} (\bibinfo {year} {2023})}\BibitemShut {NoStop}%
\bibitem [{\citenamefont {Artini}\ and\ \citenamefont {Paternostro}(2023)}]{artini2023characterizing}%
  \BibitemOpen
  \bibfield  {author} {\bibinfo {author} {\bibfnamefont {S.}~\bibnamefont {Artini}}\ and\ \bibinfo {author} {\bibfnamefont {M.}~\bibnamefont {Paternostro}},\ }\bibfield  {title} {\bibinfo {title} {Characterizing the spontaneous collapse of a wavefunction through entropy production},\ }\href {https://doi.org/10.1088/1367-2630/ad153a} {\bibfield  {journal} {\bibinfo  {journal} {New Journal of Physics}\ }\textbf {\bibinfo {volume} {25}},\ \bibinfo {pages} {123047} (\bibinfo {year} {2023})}\BibitemShut {NoStop}%
\bibitem [{\citenamefont {Ghirardi}\ \emph {et~al.}(1990)\citenamefont {Ghirardi}, \citenamefont {Pearle},\ and\ \citenamefont {Rimini}}]{ghirardi1990markov}%
  \BibitemOpen
  \bibfield  {author} {\bibinfo {author} {\bibfnamefont {G.~C.}\ \bibnamefont {Ghirardi}}, \bibinfo {author} {\bibfnamefont {P.}~\bibnamefont {Pearle}},\ and\ \bibinfo {author} {\bibfnamefont {A.}~\bibnamefont {Rimini}},\ }\bibfield  {title} {\bibinfo {title} {Markov processes in hilbert space and continuous spontaneous localization of systems of identical particles},\ }\href {https://link.aps.org/doi/10.1103/PhysRevA.42.78} {\bibfield  {journal} {\bibinfo  {journal} {Physical Review A}\ }\textbf {\bibinfo {volume} {42}},\ \bibinfo {pages} {78} (\bibinfo {year} {1990})}\BibitemShut {NoStop}%
\bibitem [{\citenamefont {Landi}\ and\ \citenamefont {Paternostro}(2021)}]{landi2021irreversible}%
  \BibitemOpen
  \bibfield  {author} {\bibinfo {author} {\bibfnamefont {G.~T.}\ \bibnamefont {Landi}}\ and\ \bibinfo {author} {\bibfnamefont {M.}~\bibnamefont {Paternostro}},\ }\bibfield  {title} {\bibinfo {title} {Irreversible entropy production: From classical to quantum},\ }\href {https://link.aps.org/doi/10.1103/RevModPhys.93.035008} {\bibfield  {journal} {\bibinfo  {journal} {Reviews of Modern Physics}\ }\textbf {\bibinfo {volume} {93}},\ \bibinfo {pages} {035008} (\bibinfo {year} {2021})}\BibitemShut {NoStop}%
\bibitem [{\citenamefont {Santos}\ \emph {et~al.}(2018{\natexlab{a}})\citenamefont {Santos}, \citenamefont {de~Paula~Jr}, \citenamefont {Drumond}, \citenamefont {Landi},\ and\ \citenamefont {Paternostro}}]{santos2018irreversibility}%
  \BibitemOpen
  \bibfield  {author} {\bibinfo {author} {\bibfnamefont {J.~P.}\ \bibnamefont {Santos}}, \bibinfo {author} {\bibfnamefont {A.~L.}\ \bibnamefont {de~Paula~Jr}}, \bibinfo {author} {\bibfnamefont {R.}~\bibnamefont {Drumond}}, \bibinfo {author} {\bibfnamefont {G.~T.}\ \bibnamefont {Landi}},\ and\ \bibinfo {author} {\bibfnamefont {M.}~\bibnamefont {Paternostro}},\ }\bibfield  {title} {\bibinfo {title} {Irreversibility at zero temperature from the perspective of the environment},\ }\href {https://link.aps.org/doi/10.1103/PhysRevA.97.050101} {\bibfield  {journal} {\bibinfo  {journal} {Physical Review A}\ }\textbf {\bibinfo {volume} {97}},\ \bibinfo {pages} {050101} (\bibinfo {year} {2018}{\natexlab{a}})}\BibitemShut {NoStop}%
\bibitem [{\citenamefont {Deffner}\ and\ \citenamefont {Lutz}(2011)}]{deffner2011nonequilibrium}%
  \BibitemOpen
  \bibfield  {author} {\bibinfo {author} {\bibfnamefont {S.}~\bibnamefont {Deffner}}\ and\ \bibinfo {author} {\bibfnamefont {E.}~\bibnamefont {Lutz}},\ }\bibfield  {title} {\bibinfo {title} {Nonequilibrium entropy production for open quantum systems},\ }\href {https://link.aps.org/doi/10.1103/PhysRevLett.107.140404} {\bibfield  {journal} {\bibinfo  {journal} {Physical Review Letters}\ }\textbf {\bibinfo {volume} {107}},\ \bibinfo {pages} {140404} (\bibinfo {year} {2011})}\BibitemShut {NoStop}%
\bibitem [{\citenamefont {Man'ko}\ \emph {et~al.}(2002)\citenamefont {Man'ko}, \citenamefont {Man'ko},\ and\ \citenamefont {Marmo}}]{man2002alternative}%
  \BibitemOpen
  \bibfield  {author} {\bibinfo {author} {\bibfnamefont {O.~V.}\ \bibnamefont {Man'ko}}, \bibinfo {author} {\bibfnamefont {V.}~\bibnamefont {Man'ko}},\ and\ \bibinfo {author} {\bibfnamefont {G.}~\bibnamefont {Marmo}},\ }\bibfield  {title} {\bibinfo {title} {Alternative commutation relations, star products and tomography},\ }\href {https://doi.org/10.1088/0305-4470/35/3/315} {\bibfield  {journal} {\bibinfo  {journal} {Journal of Physics A: Mathematical and General}\ }\textbf {\bibinfo {volume} {35}},\ \bibinfo {pages} {699} (\bibinfo {year} {2002})}\BibitemShut {NoStop}%
\bibitem [{\citenamefont {Santos}\ \emph {et~al.}(2017)\citenamefont {Santos}, \citenamefont {Landi},\ and\ \citenamefont {Paternostro}}]{santos2017wigner}%
  \BibitemOpen
  \bibfield  {author} {\bibinfo {author} {\bibfnamefont {J.~P.}\ \bibnamefont {Santos}}, \bibinfo {author} {\bibfnamefont {G.~T.}\ \bibnamefont {Landi}},\ and\ \bibinfo {author} {\bibfnamefont {M.}~\bibnamefont {Paternostro}},\ }\bibfield  {title} {\bibinfo {title} {Wigner entropy production rate},\ }\href {https://doi.org/10.1103/PhysRevLett.118.220601} {\bibfield  {journal} {\bibinfo  {journal} {Physical review letters}\ }\textbf {\bibinfo {volume} {118}},\ \bibinfo {pages} {220601} (\bibinfo {year} {2017})}\BibitemShut {NoStop}%
\bibitem [{\citenamefont {Donadi}\ and\ \citenamefont {Fadel}(2025)}]{PhysRevD.111.026009}%
  \BibitemOpen
  \bibfield  {author} {\bibinfo {author} {\bibfnamefont {S.}~\bibnamefont {Donadi}}\ and\ \bibinfo {author} {\bibfnamefont {M.}~\bibnamefont {Fadel}},\ }\bibfield  {title} {\bibinfo {title} {Quantum gravitational decoherence of a mechanical oscillator from spacetime fluctuations},\ }\href {https://doi.org/10.1103/PhysRevD.111.026009} {\bibfield  {journal} {\bibinfo  {journal} {Phys. Rev. D}\ }\textbf {\bibinfo {volume} {111}},\ \bibinfo {pages} {026009} (\bibinfo {year} {2025})}\BibitemShut {NoStop}%
\bibitem [{\citenamefont {Konopik}\ and\ \citenamefont {Lutz}(2019)}]{Konopik2019}%
  \BibitemOpen
  \bibfield  {author} {\bibinfo {author} {\bibfnamefont {M.}~\bibnamefont {Konopik}}\ and\ \bibinfo {author} {\bibfnamefont {E.}~\bibnamefont {Lutz}},\ }\bibfield  {title} {\bibinfo {title} {Quantum response theory for nonequilibrium steady states},\ }\href {https://doi.org/https://doi.org/10.1103/PhysRevResearch.1.033156} {\bibfield  {journal} {\bibinfo  {journal} {Phys. Rev. Research}\ }\textbf {\bibinfo {volume} {1}},\ \bibinfo {pages} {033156} (\bibinfo {year} {2019})}\BibitemShut {NoStop}%
\bibitem [{\citenamefont {Blair}\ \emph {et~al.}(2024)\citenamefont {Blair}, \citenamefont {Zicari}, \citenamefont {Belenchia}, \citenamefont {Ferraro},\ and\ \citenamefont {Paternostro}}]{Blair2024}%
  \BibitemOpen
  \bibfield  {author} {\bibinfo {author} {\bibfnamefont {S.}~\bibnamefont {Blair}}, \bibinfo {author} {\bibfnamefont {G.}~\bibnamefont {Zicari}}, \bibinfo {author} {\bibfnamefont {A.}~\bibnamefont {Belenchia}}, \bibinfo {author} {\bibfnamefont {A.}~\bibnamefont {Ferraro}},\ and\ \bibinfo {author} {\bibfnamefont {M.}~\bibnamefont {Paternostro}},\ }\bibfield  {title} {\bibinfo {title} {Non-equilibrium quantum probing through linear response},\ }\href {https://doi.org/https://doi.org/10.1103/PhysRevResearch.6.013152} {\bibfield  {journal} {\bibinfo  {journal} {Phys. Rev. Research}\ }\textbf {\bibinfo {volume} {6}},\ \bibinfo {pages} {013152} (\bibinfo {year} {2024})}\BibitemShut {NoStop}%
\bibitem [{\citenamefont {Uzdin}\ and\ \citenamefont {Rahav}(2021)}]{uzdin2021passivity}%
  \BibitemOpen
  \bibfield  {author} {\bibinfo {author} {\bibfnamefont {R.}~\bibnamefont {Uzdin}}\ and\ \bibinfo {author} {\bibfnamefont {S.}~\bibnamefont {Rahav}},\ }\bibfield  {title} {\bibinfo {title} {Passivity deformation approach for the thermodynamics of isolated quantum setups},\ }\href {https://doi.org/10.1103/PRXQuantum.2.010336} {\bibfield  {journal} {\bibinfo  {journal} {PRX Quantum}\ }\textbf {\bibinfo {volume} {2}},\ \bibinfo {pages} {010336} (\bibinfo {year} {2021})}\BibitemShut {NoStop}%
\bibitem [{\citenamefont {Cover}\ and\ \citenamefont {Thomas}(1991)}]{cover1991information}%
  \BibitemOpen
  \bibfield  {author} {\bibinfo {author} {\bibfnamefont {T.~M.}\ \bibnamefont {Cover}}\ and\ \bibinfo {author} {\bibfnamefont {J.~A.}\ \bibnamefont {Thomas}},\ }\bibfield  {title} {\bibinfo {title} {Information theory and statistics},\ }\href@noop {} {\bibfield  {journal} {\bibinfo  {journal} {Elements of information theory}\ }\textbf {\bibinfo {volume} {1}},\ \bibinfo {pages} {279} (\bibinfo {year} {1991})}\BibitemShut {NoStop}%
\bibitem [{\citenamefont {Di{\'o}si}(2007)}]{diosi2007notes}%
  \BibitemOpen
  \bibfield  {author} {\bibinfo {author} {\bibfnamefont {L.}~\bibnamefont {Di{\'o}si}},\ }\bibfield  {title} {\bibinfo {title} {Notes on certain {N}ewton gravity mechanisms of wavefunction localization and decoherence},\ }\href {https://doi.org/10.1088/1751-8113/40/12/S07} {\bibfield  {journal} {\bibinfo  {journal} {Journal of Physics A: Mathematical and Theoretical}\ }\textbf {\bibinfo {volume} {40}},\ \bibinfo {pages} {2989} (\bibinfo {year} {2007})}\BibitemShut {NoStop}%
\bibitem [{\citenamefont {Figurato}\ \emph {et~al.}(2024)\citenamefont {Figurato}, \citenamefont {Dirindin}, \citenamefont {Gaona-Reyes}, \citenamefont {Carlesso}, \citenamefont {Bassi},\ and\ \citenamefont {Donadi}}]{figurato2024effectiveness}%
  \BibitemOpen
  \bibfield  {author} {\bibinfo {author} {\bibfnamefont {L.}~\bibnamefont {Figurato}}, \bibinfo {author} {\bibfnamefont {M.}~\bibnamefont {Dirindin}}, \bibinfo {author} {\bibfnamefont {J.~L.}\ \bibnamefont {Gaona-Reyes}}, \bibinfo {author} {\bibfnamefont {M.}~\bibnamefont {Carlesso}}, \bibinfo {author} {\bibfnamefont {A.}~\bibnamefont {Bassi}},\ and\ \bibinfo {author} {\bibfnamefont {S.}~\bibnamefont {Donadi}},\ }\bibfield  {title} {\bibinfo {title} {On the effectiveness of the collapse in the di{\'o}si--penrose model},\ }\href {https://doi.org/https://doi.org/10.1088/1367-2630/ad8c77} {\bibfield  {journal} {\bibinfo  {journal} {New Journal of Physics}\ }\textbf {\bibinfo {volume} {26}},\ \bibinfo {pages} {113004} (\bibinfo {year} {2024})}\BibitemShut {NoStop}%
\bibitem [{\citenamefont {Baker~Jr}(1958)}]{baker1958formulation}%
  \BibitemOpen
  \bibfield  {author} {\bibinfo {author} {\bibfnamefont {G.~A.}\ \bibnamefont {Baker~Jr}},\ }\bibfield  {title} {\bibinfo {title} {Formulation of quantum mechanics based on the quasi-probability distribution induced on phase space},\ }\href {https://doi.org/https://doi.org/10.1103/PhysRev.109.2198} {\bibfield  {journal} {\bibinfo  {journal} {Physical Review}\ }\textbf {\bibinfo {volume} {109}},\ \bibinfo {pages} {2198} (\bibinfo {year} {1958})}\BibitemShut {NoStop}%
\bibitem [{\citenamefont {Bahrami}\ \emph {et~al.}(2014)\citenamefont {Bahrami}, \citenamefont {Smirne},\ and\ \citenamefont {Bassi}}]{bahrami2014role}%
  \BibitemOpen
  \bibfield  {author} {\bibinfo {author} {\bibfnamefont {M.}~\bibnamefont {Bahrami}}, \bibinfo {author} {\bibfnamefont {A.}~\bibnamefont {Smirne}},\ and\ \bibinfo {author} {\bibfnamefont {A.}~\bibnamefont {Bassi}},\ }\bibfield  {title} {\bibinfo {title} {Role of gravity in the collapse of a wave function: A probe into the {D}i{\'o}si-{P}enrose model},\ }\href {https://doi.org/https://doi.org/10.1103/PhysRevA.90.062105} {\bibfield  {journal} {\bibinfo  {journal} {Physical Review A}\ }\textbf {\bibinfo {volume} {90}},\ \bibinfo {pages} {062105} (\bibinfo {year} {2014})}\BibitemShut {NoStop}%
\bibitem [{Note1()}]{Note1}%
  \BibitemOpen
  \bibinfo {note} {Consider the Shannon entropy of a factorized probability distribution: $p(x_1,x_2,x_3)=p_1(x_1)p_2(x_2)p_3(x_3)$, it is easy to show that $S(p) = S(p_1)+S(p_2)+S(p_3)$. The same holds for the relative entropy as long as the reference distribution is factorized as well.}\BibitemShut {Stop}%
\bibitem [{\citenamefont {Swift}\ and\ \citenamefont {Hohenberg}(1977)}]{swift1977hydrodynamic}%
  \BibitemOpen
  \bibfield  {author} {\bibinfo {author} {\bibfnamefont {J.}~\bibnamefont {Swift}}\ and\ \bibinfo {author} {\bibfnamefont {P.~C.}\ \bibnamefont {Hohenberg}},\ }\bibfield  {title} {\bibinfo {title} {Hydrodynamic fluctuations at the convective instability},\ }\href {https://doi.org/https://doi.org/10.1103/PhysRevA.15.319} {\bibfield  {journal} {\bibinfo  {journal} {Physical Review A}\ }\textbf {\bibinfo {volume} {15}},\ \bibinfo {pages} {319} (\bibinfo {year} {1977})}\BibitemShut {NoStop}%
\bibitem [{\citenamefont {Kramers}(1940)}]{kramers1940brownian}%
  \BibitemOpen
  \bibfield  {author} {\bibinfo {author} {\bibfnamefont {H.~A.}\ \bibnamefont {Kramers}},\ }\bibfield  {title} {\bibinfo {title} {Brownian motion in a field of force and the diffusion model of chemical reactions},\ }\href {https://doi.org/https://doi.org/10.1016/S0031-8914(40)90098-2} {\bibfield  {journal} {\bibinfo  {journal} {{P}hysica}\ }\textbf {\bibinfo {volume} {7}},\ \bibinfo {pages} {284} (\bibinfo {year} {1940})}\BibitemShut {NoStop}%
\bibitem [{\citenamefont {Tom{\'e}}\ and\ \citenamefont {de~Oliveira}(2010)}]{tome2010entropy}%
  \BibitemOpen
  \bibfield  {author} {\bibinfo {author} {\bibfnamefont {T.}~\bibnamefont {Tom{\'e}}}\ and\ \bibinfo {author} {\bibfnamefont {M.~J.}\ \bibnamefont {de~Oliveira}},\ }\bibfield  {title} {\bibinfo {title} {Entropy production in irreversible systems described by a fokker-planck equation},\ }\href@noop {} {\bibfield  {journal} {\bibinfo  {journal} {Physical Review E—Statistical, Nonlinear, and Soft Matter Physics}\ }\textbf {\bibinfo {volume} {82}},\ \bibinfo {pages} {021120} (\bibinfo {year} {2010})}\BibitemShut {NoStop}%
\bibitem [{\citenamefont {Santos}\ \emph {et~al.}(2018{\natexlab{b}})\citenamefont {Santos}, \citenamefont {Céleri}, \citenamefont {Brito}, \citenamefont {Landi},\ and\ \citenamefont {Paternostro}}]{Santos2018}%
  \BibitemOpen
  \bibfield  {author} {\bibinfo {author} {\bibfnamefont {J.~P.}\ \bibnamefont {Santos}}, \bibinfo {author} {\bibfnamefont {L.~C.}\ \bibnamefont {Céleri}}, \bibinfo {author} {\bibfnamefont {F.}~\bibnamefont {Brito}}, \bibinfo {author} {\bibfnamefont {G.~T.}\ \bibnamefont {Landi}},\ and\ \bibinfo {author} {\bibfnamefont {M.}~\bibnamefont {Paternostro}},\ }\bibfield  {title} {\bibinfo {title} {Spin-phase-space-entropy production},\ }\href {http://dx.doi.org/10.1103/PhysRevA.97.052123} {\bibfield  {journal} {\bibinfo  {journal} {Physical Review A}\ }\textbf {\bibinfo {volume} {97}},\ \bibinfo {pages} {052123} (\bibinfo {year} {2018}{\natexlab{b}})}\BibitemShut {NoStop}%
\bibitem [{Note2()}]{Note2}%
  \BibitemOpen
  \bibinfo {note} {The dissipative DP model is not well-defined in one dimension because the $\Gamma (k)$ term, which is proportional to $1/k^2$ [cf.~\protect \cref {eq: Gamma}], can be integrated in three dimensions, but leads to divergences in one dimension, much like the Newtonian potential.}\BibitemShut {Stop}%
\bibitem [{\citenamefont {Navarrete-Benlloch}(2022)}]{navarrete2022introduction}%
  \BibitemOpen
  \bibfield  {author} {\bibinfo {author} {\bibfnamefont {C.}~\bibnamefont {Navarrete-Benlloch}},\ }\bibfield  {title} {\bibinfo {title} {Introduction to quantum optics},\ }\href@noop {} {\bibfield  {journal} {\bibinfo  {journal} {arXiv preprint arXiv:2203.13206}\ } (\bibinfo {year} {2022})}\BibitemShut {NoStop}%
\end{thebibliography}%

\end{document}